\documentclass[11pt,a4paper]{article}
\usepackage[inactive, tightpage]{preview}
\usepackage{jheppub}
\usepackage{amsthm}
\usepackage[T1]{fontenc}
\usepackage[utf8]{inputenc}
\usepackage{mathtools}   
\usepackage{amsfonts}
\usepackage{physics}
\usepackage{mathrsfs}
\usepackage{natbib}
\usepackage{comment}
\usepackage{ytableau}
\usepackage{tikz, pgfplots}
\usepackage{ytableau}
\usetikzlibrary{patterns, calc, patterns, angles, quotes, decorations, shadows}
\usepackage{enumitem}
\usepackage{subcaption}
\usepackage[font={small}]{caption}
\usepackage{graphicx}
\usepackage{amssymb}
\usepackage{microtype}
\usepackage{todonotes}

\usepackage{microtype}
\usepackage{hyperref}

\definecolor{darkmidnightblue}{rgb}{0.0, 0.2, 0.4}
\hypersetup{
    colorlinks = true,
    linkcolor=darkmidnightblue,
    citecolor=darkmidnightblue
}

\usepackage{cleveref}

\usepackage{booktabs}
\usepackage{pdflscape}
\usepackage{afterpage}
\usepackage{flafter}
\usepackage{rotating}
\usepackage{fancyhdr}
\usepackage{slashed}
\fancypagestyle{lscape}{%
\fancyhf{} 
\fancyfoot[LE]{}
\fancyfoot[LO] {}

}

\newcommand{\Z}{\mathbb{Z}}
\newcommand{\R}{\mathbb{R}} 
\newcommand{\C}{\mathbb{C}} 
\newcommand{\ch}{\text{ch}}
\DeclareMathSymbol{:}{\mathord}{operators}{"3A}

\newcommand{\blue}[1]{{\color{blue} #1 \color{black}}}

\newcommand{\beq}{\begin{eqnarray}}
\newcommand{\eeq}{\end{eqnarray}}
\newcommand{\bea}{\begin{eqnarray}}
\newcommand{\eea}{\end{eqnarray}}
\newcommand{\be}{\begin{equation}}
\newcommand{\ee}{\end{equation}}
\newcommand{\bq}{\begin{equation}}
\newcommand{\eq}{\end{equation}}

\newcommand{\half}{\frac{1}{2}}
\newcommand{\nn}{\nonumber}

\newcommand{\PB}[1]{{\blue{PB: #1}}} 

\title{Factorizations of 3d Interval Partition Functions}

\author[a]{Boan Zhao}
\author[b]{Panos Betzios}
\author[a]{Paul Luis Roehl}

\affiliation[a]{DAMTP, University of Cambridge, Cambridge, United Kingdom}

\affiliation[b]{\href{https://www.ugent.be/we/physics-astronomy/en}{Department of Physics and Astronomy},
Ghent University, \\ Krijgslaan, 281-S9, 9000 Gent, Belgium}

\emailAdd{bz258@cam.ac.uk}
\emailAdd{panos.betzios@ugent.be}
\emailAdd{plr30@cam.ac.uk}

\abstract{
We show that interval partition functions (transition amplitudes) of three-dimensional $\mathcal{N} = 2$ theories admit factorizations into sums of products of hemisphere partition functions with Wilson loop insertions glued by suitable factors. We prove the factorization explicitly for supersymmetric quantum electrodynamics and Chern-Simons-Yang-Mills theories. In the former case, we show that the gluing factors can be naturally interpreted in terms of $S^2 \times S^1$ partition functions. In the latter case, we prove that hemisphere partition functions are affine characters and determine the gluing factors explicitly in special cases.
}

\theoremstyle{definition}

\begin{document}
\ytableausetup{boxsize=0.5em}
\ytableausetup{centertableaux}
\maketitle

\section{Introduction}
\begin{figure}[t]
	\centering
	\includegraphics{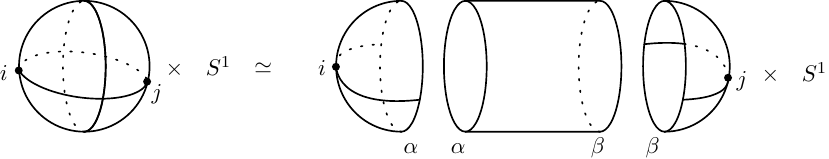}
	\caption{Factorization of the $S^2 \times S^1$ partition function into hemiphere partition functions glued by interval partition functions. The $i,j$ denote Wilson loops wrapping the $S^1$. }
    \label{fig: sphere}
\end{figure}

Supersymmetric partition functions have been known to obey interesting mathematical relations, one of which is the factorization of the topologically twisted $S^2 \times S^1$ partition functions into twisted hemisphere \footnote{$HS^2 \times S^1$ is sometimes also known as the solid torus.} partitions on $HS^2 \times S^1$ written schematically as:
\begin{eqnarray}
	S = \sum_{\alpha, \beta}H_{\alpha} I^{\alpha\beta}\tilde{H}_{\beta}
\end{eqnarray}
where $S$ stands for an $S^2\times S^1$ partition function (without any Wilson loop insertion) and $H_\alpha$ are partition functions on $HS^2 \times S^1$ with the boundary condition labelled by $\alpha$. $I^{\alpha\beta}$ is a suitable gluing factor. Geometrically, it corresponds to decomposing $S^2 \times S^1$ into the union of two $HS^2 \times S^1$. We then insert a basis of boundary states $\ket{B_\alpha},\ket{B_\beta}$ at the boundary tori. The geometric picture (figure  \ref{fig: sphere}) \footnote{For now we assume that there is no Wilson loop insertion $i = j = 0$. Later we will work with the factorization with general $i,j$.} shows that $I^{\alpha\beta}$ is the inverse of the inner product matrix $\braket{B_\alpha}{B_\beta}$. The inner product $\braket{B_\alpha}{B_\beta}$ equals the interval partition function $I_{\alpha\beta}$ on $T^2 \times [0, 1]$  with boundary conditions $\alpha, \beta$ at the two boundary tori.

\begin{figure}[t]
	\centering
	\includegraphics{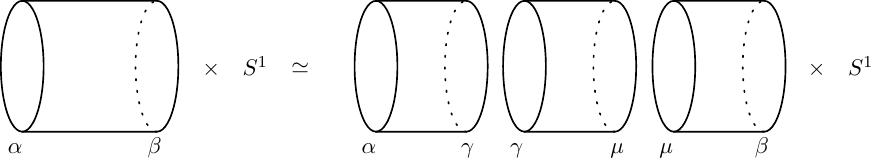}
    
	\caption{Factorization of interval partition functions into interval partition functions by inserting a complete basis of states $\gamma, \mu$.}
    \label{fig: int_fact_int}
\end{figure}

Factorizations of supersymmetric partition functions on manifolds with boundary have received less attention. One example is the factorization of the 3d interval partition functions on $T^2 \times [0, 1]$ into interval partition functions \cite{ZhangThesis}. Let $I_{\alpha\beta}$ denote the interval partition function with boundary conditions $\alpha,\beta$ at the two boundary tori. The factorization is:
\begin{eqnarray}
	I_{\alpha\beta} = \sum_{\gamma, \mu}I_{\alpha \gamma} I^{\gamma \mu}          I_{\mu \beta}
\end{eqnarray}
Geometrically, it corresponds to splitting the interval into two intervals and inserting a basis of states in the middle (figure \ref{fig: int_fact_int}).

In this work, we study a different type of factorization of interval partition functions. We aim to factorize them into sums of products of hemisphere partition functions with suitable gluing factors. Written schematically, it is
\begin{eqnarray}\label{eq: fact_int}
	I_{\alpha\beta} = \sum_{i, j} H_{\alpha i} S^{ij} \tilde{H}_{\beta j}
\end{eqnarray}
where $i, j$ label Wilson loops inserted at the tip of the hemisphere. The hemisphere partition function $H$ now depends on the boundary condition $\alpha$ at the boundary torus and the Wilson loop $i$ inserted at the tip wrapping the $S^1$. $\tilde{H}$ is another hemisphere partition function related to $H$ by a suitable transformation of variables. $S^{ij}$ is a suitable gluing factor.  We shall see that in some cases $S^{ij}$ is the inverse of the $S^2 \times S^1$ partition functions with Wilson loops $i,j$ inserted at the two poles as illustrated by the following picture \ref{fig: int_fact}.
\begin{figure}[t]
\label{fig: int_fact}
	\centering
	\includegraphics{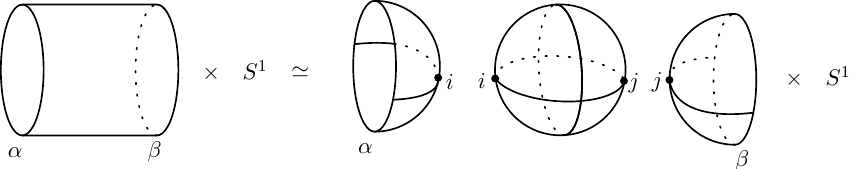}
	\caption{Factorization of interval partition functions into hemisphere partition functions by inserting a complete basis of Wilson loops $i, j$ at the tips of the hemispheres. $\alpha, \beta$ are boundary conditions at the boundary tori.}
    \label{fig: int_fact}
\end{figure}
We now briefly describe the two examples in the main text to illustrate \eqref{eq: fact_int}.

Our first example consists of a 3d $\mathcal{N} = 2$ $U(1)$ vector multiplet plus $N$ charge 1 chirals $X_1,...,X_N$ with flavour symmetry fugacities $a_1,..., a_N$. We assume that $N$ is even to avoid parity anomalies. We use the following basis of boundary conditions \footnote{The boundary conditions for the gauge field will be explained in the main text. Other choices of boundary conditions are possible. We chose this set of boundary conditions to illustrate the method.}  $\alpha = 1,..., N$:
\begin{eqnarray}
	B_\alpha:\quad X_\alpha = c\neq 0\quad X_\beta = 0, \forall \beta \neq \alpha
\end{eqnarray}
where $c$ is a nonzero constant. In other words, only the scalar $X_\alpha$ acquires a nonzero VEV on the boundary and every other scalar vanishes on the boundary. The gauge symmetry is completely broken on the boundary so we do not require any Chern-Simons level/boundary matter to cancel any boundary gauge anomaly. The Wilson loops inserted at the tip of the hemisphere are labelled by $i\in \Z$ corresponds to the $i$th power of the fundamental representation of $U(1)$ \footnote{For example $i = 2$ correspond to the representation $z\to z^2$ where $z\in U(1)$}. In this case, the factorization of the interval partition function becomes
\begin{eqnarray}
	I_{\alpha\beta}(q, a) = \sum_{i, j = 1}^N H_{\alpha i}(\Lambda, q^{-1}, a)S^{ij}(\Lambda, q , a)H_{\beta j}(\Lambda, q, a)
\end{eqnarray}
where $S^{ij}$ is the inverse of the $S^2\times S^1$ partition function with Wilson loops $i,j$ inserted at the two poles wrapping the $S^1$. $\Lambda$ is the vortex grading parameter. We prove this factorization by inverting the factorization of $S^2\times S^1$ partition function
\begin{eqnarray}
	S_{ij}(\Lambda, q, a)=  \sum_{\alpha = 1}^N H_{\alpha i}(\Lambda,q, a)H_{\alpha j}(\Lambda, q^{-1}, a)I_{\alpha\alpha}(q, a)^{-1}
\end{eqnarray}
We prove that $S_{ij}$ is invertible as a matrix when $\Lambda$ is sufficiently small.

Our next example is the interval partition functions of 3d $\mathcal{N} = 2$ $U(k)$ Chern-Simons-Yang-Mills theory coupled to $N$ boundary fundamental Fermi multiplets on each boundary torus. The Chern-Simons level is set to $-k - N$ to cancel the boundary gauge anomaly. We use Neumann boundary conditions on the gauge fields. The boundary Fermi multiplets have $SU(N) \times SU(N)$ flavour symmetry and we use fugacities $x_1,...,x_N$ for the first $SU(N)$ and $y_1,...,y_N$ for the second $SU(N)$. We will show that the interval partition function admits a diagonal factorization into affine characters:

\begin{eqnarray}
	I^{N, k}(q, x, y) = \sum_{\mu} f^{N, k}_{\mu\mu}(q) \chi_{\mu}^{\widehat{SU(N)}_k}(q, x) \chi_{\mu}^{\widehat{SU(N)}_k}(q, y)
\end{eqnarray}
where $ \chi_{\mu}^{\widehat{SU(N)}_k}$ are characters of affine $\widehat{SU(N)}$ at level $k$ with highest weight $\mu$ (appendix \ref{app:su(n)}). We also prove that the hemisphere partition functions with Wilson loop $\mu$ inserted at the tip equal $\chi_{\mu}^{\widehat{SU(N)}_k}$ up to a prefactor of $q$. Therefore, the factorization can be regarded as a factorization into hemisphere partition functions by suitable redefinitions of the functions $f_{\mu\mu}^{N,k}(q)$. The physical interpretation \footnote{Our argument for the $U(1)$ gauge theory does not immediately apply here due to a mismatch between the fermion periodicity along the hemisphere and one of the cycles of the torus. This will be elaborated upon near the start of section \ref{sec: cs_ym}.} is that under the level-rank duality, this system becomes $\mathcal{N} = 2$ $SU(N)_{k + N}$  Chern-Simons-Yang-Mills on the interval which flows to $SU(N)_k$ pure Chern-Simons theory in the infrared. Therefore, we expect a diagonal factorization into affine characters via the Chern-Simons/WZW duality. The physical interpretations of the gluing factors $f_{\mu\mu}^{N,k}$ will be left for future work.

We expect our analysis to generalize in a straightforward manner to other supersymmetric setups across other dimensions.

\section{$U(1)$ vector multiplet with $N$ unit charge chirals}
In this section we factorize the interval partition functions of the 3d $\mathcal{N} = 2$ $U(1)$ vector multiplet with $N$  chirals with unit gauge charge. We denote the dynamical bosons as $A_\mu, \sigma, X_1,..., X_N$, where $(A_\mu, \sigma )$ lie in the vector multiplet and $X_1,...,X_N$ lie in the $N$ chirals respectively. The Lagrangian for this system can be found in \cite{ClossetReview}.

We place this system on the following three geometries:
\begin{eqnarray}
	S^2 \times S^1_q  \quad HS^2 \times S^1_q \quad S^1 \times [0,1] \times S^1_q
\end{eqnarray}
where $HS^2$ is the two-dimensional hemisphere and $S^1 \times [0,1]$ is differomorphic to the cylinder. The last geometry is also diffeomorphic to $T^2 \times [0, 1]$ will be referred to as the cylinder/interval geometry. The $S^1_q$ factor is the time circle and has length $\beta$ \footnote{The supersymmetric partition functions depends on $\beta$ only through the background holonomies $a_1,..., a_N$}.  We use $x^0$ as the coordinate along $S^1_q$ and $z$ as the complex coordinates along $S^2, HS^2, S^1 \times [0, 1]$. Therefore, the components of the gauge fields (both dynamical and background) can be written as $A_0, A_z, A_{\bar{z}}$. We perform a topological twist along the spatial directions  \cite{ZhaoBLT, ZhaoThesis} and preserve two scalar supercharges.

The subscript  $q\in U(1)$ in $S^1_q$ indicates that we switch on an $\Omega$ -background along this direction. We assume that $q$ is generic. If $S^1_q$ has length $\beta$ then for a generic field $\Phi$:
\begin{eqnarray}
	\Phi(x^0 = \beta) = q^{-J} \Phi(x^0 = 0)
\end{eqnarray}
where $J$ is the rotation operator on $S^2, HS^2, S^1 \times [0,1]$ \footnote{All three geometries carry natural $U(1)$ isometric actions.}. In other words, $\Phi$ restricted to the time slice $x^0 = \beta$  is a rotated version of $\Phi$ restricted to $x^0 = 0$, rotated by the angle $\log(q)/i$. Since $\Phi$ lies in a (possibly topologically non-trivial) line bundle over the spacetime, we need to pick an uplift of the $J$ operator to the total space of the line bundle. The physical partition function does not depend on this uplift.

This system has a $U(1)^N$ flavour symmetry $(X_1,..., X_N)\to (a_1X_1,..., a_NX_N), a_i\in U(1)$  which allows us to switch on constant background $U(1)$ gauge fields $A_0^k, k =1,...,N$ along $S^1_q$ with holonomies $a_k = \exp(-i\beta A_0^k)$ which couple to the matter scalars via the flavour symmetry action. Therefore, the partition functions on all three geometries are functions of $q, a_1,...,a_N$. We will use the notation $a = (a_1,...,a_N)$ throughout this section.

We can then insert the following BPS Wilson loops (see eqn. 4.7 of \cite{Benini3d}). 
\begin{eqnarray}\label{eq: wilson_loop}
	W_k = \exp(\int_{S^1_q} (-iA_0 + \sigma)dx^0)^k q^{k\text{Weight} (\text{Fibre})} \, ,
\end{eqnarray}
at the torus fixed points of the two geometries $S^2, HS^2$  and wrapping $S^1_q$. The notation $\text{Weight(Fibre)}$ denotes the weight (an integer) of the lift of the $U(1)$ rotation of $S^2$ or $ HS^2$ to the fibre of the gauge line bundle above the fixed point.  This term is needed so that the expectation of the Wilson loop is independent of the choice of the uplift of the $U(1)$ spacetime rotation.  On $S^2$ we have a Wilson loop $W_k$ at the north pole and $W_j$ at the south pole. The resulting partition function is denoted as $S_{ij} (q, a)$ where $a = (a_1,...,a_N)$ and $i,j\in \Z$. Later we will restrict $i,j = 0,..., N - 1$. On the hemisphere we only have one Wilson loop $W_k$ inserted at the tip of the hemisphere. 

We use the supersymmetric completions of the following set of boundary conditions $\alpha = 1,..., N$ \cite{ZhaoBLT, ZhaoThesis}
\begin{equation}
	\begin{aligned}
		B_\alpha:\quad &X_\alpha  = c\neq 0\quad X_\beta = 0 \quad\forall \beta \neq \alpha \\
		&\nabla_zA_{\bar{z}} = 0\quad A_0 + i\sigma = const
	\end{aligned}
\end{equation}
where $c$ is a fixed nonzero constant. We integrate over all possible constant values of $A_0 + i\sigma$ in the path integral
for $HS^2 \times S^1$. Therefore, the hemisphere partition function is denoted $H_{k\alpha}(q, x)$ where $k$ stands for the Wilson loop $W_k$ at the tip of the hemisphere and $\alpha$ denotes the boundary condition. On the cylinder/interval geometry $S^1 \times [0,1] \times S^1_q$, we can choose independent boundary conditions $B_\alpha, B_\beta$ at the two boundary tori. The interval partition function is denoted $I_{\alpha\beta}(q, a)$ where $\alpha,\beta$ denote the two boundary conditions.

A crucial feature of the boundary condition $X_\alpha =c$ is that we need to trivialize the gauge bundle on the boundary torus \cite{DavideDualBC}. The trivialization of the gauge bundle on the boundary circle of $HS^2$ is classified topologically by its winding number $d$  relative to a trivialization of the gauge bundle on the whole $HS^2$. This is the vortex number which we will discuss in the next section. Indeed, in a trivialisation of the whole gauge bundle on $HS^2$, $d$ is the winding number of the phase of $X_\alpha$ on the boundary circle. On the interval/cylinder geometry, we need to classify pairs of boundary trivializations up to automorphisms of the whole gauge bundle on $S^1 \times [0,1]$. These are classified by their relative winding number \footnote{We smoothly transport the trivialization of the gauge bundle on the first boundary circle along the interval direction from $0$ to 1. Then we compute the winding number relative to the trivialization on the second boundary}. The boundary condition $\nabla_zA_{\bar{z}} = 0$ also requires a trivialization of the gauge bundle in a small neighborhood of the boundary torus. The topological classification does not change.

We choose these boundary conditions primarily as an example to illustrate the method. We expect that the factorization of the interval partition function is valid for any choice of BPS boundary conditions at the two boundary tori. The partition functions with this choice of boundary conditions can be localized to a BPS locus of finitely many points. Indeed, the $B_\alpha$ corresponds to the $\alpha$th massive vacuum on the Higgs branch if we switch on real masses. On the hemisphere, the path integral would localize to quasimaps to the massive vacuum, meaning that $X_\alpha$ has a vortex configuration on the hemisphere and all other scalars vanish at the BPS locus. On the interval, we only have a nonempty BPS locus if $\alpha = \beta$ as the scalar and the gauge fields need to be constant. We now provide more details on the computations of the partition functions on the three geometries of interest.

\subsection{Hemisphere Partition Function}
The topologically twisted hemisphere partition function $H_{\alpha i}(\Lambda, q, a)$ with Wilson loops $W_i$ inserted at the tip and the boundary condition $\alpha$  can be computed following the procedure in \cite{ZhaoBLT}. The result is
\begin{eqnarray}
	H_{\alpha i}(\Lambda, q, a) = H_{\alpha}^{\text{pert}}(q, a)H_{\alpha i}^{\text{vortex}}(\Lambda, q, a)
\end{eqnarray}
where
\begin{eqnarray}
	H_{\alpha}^{\text{pert}}(q,a ) = \prod_{n\geq 1}\prod_{ i\neq \alpha}\sqrt{q^{n} a_i/a_\alpha} - \sqrt{q^{-n} a_\alpha/a_i}\\
	H_{\alpha k}^{\text{vortex}}(\Lambda, q, a) =  \sum_{d\geq 0}\Lambda^d \frac{(q^d /a_\alpha)^k}{\prod_{n = 1}^d\prod_{i = 1}^N\sqrt{q^{n} a_i/a_\alpha} - \sqrt{q^{-n}a_\alpha/a_i}}
\end{eqnarray}

The vortex partition function converges absolutely for $|\Lambda| < 1$ and $ 0<| q|<1$. To prove this, the ratio of the coefficients at order $\Lambda^{d + 1}$ with respect to that at order $\Lambda^d$ is:
\begin{eqnarray}
    \frac{q^k}{\sqrt{q^da_i/a_\alpha} - \sqrt{q^{-d}a_\alpha/a_i}}
\end{eqnarray}
As $d$ tends to infinity, this ratio tends to zero due to the $\sqrt{q^{-d}}$ in the denominator, regardless of what $k$ is. As a result the coefficients at order $\Lambda^d$ tend to zero as $d\to \infty$.

The partition function cannot be computed using an integral over the Coulomb branch due to the boundary condition $X_\alpha \neq 0$. The Higgs branch localization (section 4.2 of \cite{ZhaoBLT})  localizes to a vortex configuration for $X_\alpha $ and $X_i = 0, \forall i \neq \alpha$. 
\begin{equation}\label{eq: vortex}
\begin{aligned}
\partial_{\bar{z}}X_\alpha - iA_{\bar{z}}X_\alpha = 0\\
	2iF_{z\bar{z}} = 1 - |X_\alpha|^2
\end{aligned}
\end{equation}
The existence and uniqueness of the vortex equation for $c = 1$ \footnote{For generic $c$, the constant term in the second vortex equation (for $F_{z\bar{z}}$) needs to be modified which amounts to tuning a constant in the localising action. We believe that the existence and uniqueness theorem of \cite{NassirVortex} still applies in that case.} can be proven following the method in \cite{ZhaoBLT, ZhaoVortex, NassirVortex, ZhaoThesis}. The one-loop determinants were computed in section 4.3.1 of \cite{ZhaoBLT}. The chiral multiplet $X_i, i\neq \alpha$ contributes
\begin{eqnarray}
	\prod _{n\geq d + 1, i\neq \alpha}\sqrt{q^{n} a_i/a_\alpha} - \sqrt{q^{-n}a_\alpha/a_i}
\end{eqnarray}
in the degree $d$ sector. The $\alpha$th chiral contributes
\begin{eqnarray}
	\prod_{n\geq d + 1}(\sqrt{q^{n}} - \sqrt{q^{-n}})
\end{eqnarray}
The vector multiplet contributes
\begin{eqnarray}
	\prod_{n\geq 1}(\sqrt{q^{n}} - \sqrt{q^{-n}})^{-1}
\end{eqnarray}
Finally, the Wilson loop contributes $(q^d/a_\alpha)^k$ \cite{ZhangThesis}. After we multiply everything together, we obtain the desired result.

$H_\alpha^{\text{pert}}(q, a)$ needs to be regularized. We can use the following regularization \cite{ZhaoBLT}:
\begin{equation}
\begin{aligned}
           H_\alpha^{\text{pert}}(q, a) = &\prod_{\beta \neq \alpha}(-1)^\infty(qa_\beta/a_\alpha; q)_\infty \prod_{n\geq 1}\sqrt{q^{-n}a_\alpha/a_\beta} \\
           = &\prod_{\beta \neq \alpha}(qa_\beta/a_\alpha;q)_\infty \exp(\frac{1}{4}\log(q^{-1}a_\alpha/a_\beta) - \frac{\log(q^{-1}a_\alpha/a_\beta)^2}{4\log(q^{-1})})
           \end{aligned}
\end{equation}
We also regularize $H_\alpha^{\text{pert}}(q^{-1}, a)$ as the following:
\begin{equation}
\begin{aligned}
           H_\alpha^{\text{pert}}(q^{-1}, a) =& \prod_{\beta \neq \alpha}(qa_\alpha/a_\beta; q)_\infty\prod_{n\geq 1} \sqrt{q^{-n}a_\beta/a_\alpha} \\
           =&\prod_{\beta \neq \alpha}(qa_\alpha/a_\beta; q)_\infty \exp(\frac{1}{4}\log(q^{-1}a_\beta/a_\alpha) - \frac{\log(q^{-1}a_\beta/a_\alpha)^2}{4\log(q^{-1})})
\end{aligned}
\end{equation}
The choice of regularization does not affect the factorization of $S^2 \times S^1$ index as long as we use a consistent regularization for both the hemisphere and the interval partition functions.

\subsection{Interval Partition Functions}
The topologically twisted interval partition function $I_{\alpha\beta}(q, a)$ with boundary conditions $\alpha, \beta$ at the two boundary tori can be computed following the procedure in \cite{ZhaoBLT}. The result is
\begin{equation}\label{eq: int_sqed}
\begin{aligned}
	I_{\alpha\beta}(q, a) &= \delta_{\alpha\beta}\prod _{n\in \Z, \gamma\neq \alpha}\sqrt{q^{n} a_\gamma/a_\alpha} - \sqrt{q^{-n}a_\alpha/a_\gamma} \\
    &= Z_\alpha^{\text{pert}}(q, a)Z_\alpha^{\text{pert}}(q^{-1}, a)\prod_{\gamma\neq \alpha}\sqrt{a_\gamma/a_\alpha} - \sqrt{a_\alpha/a_\gamma}
\end{aligned}
\end{equation}
To derive this, we again use the Higgs branch localization (section 4.2 of \cite{ZhaoBLT}). We will show that the BPS locus is a point. We again assume $c = 1$ to match the boundary condition for the hemisphere. We will show that if
\begin{enumerate}
\item[(a)] $|X_\alpha| = c$ on both boundary tori
\item[(b)] $(X_\alpha, A_\mu)$ satisfies the vortex equation \eqref{eq: vortex}
\end{enumerate}
then
\begin{enumerate}
    \item[(a)] the relative winding number of the two trivialisations at the two boundary tori must be zero
    \item[(b)] $X_\alpha$ must be gauge equivalent to a constant
\end{enumerate}

Indeed, the fact that $X_\alpha$ is rotationally invariant and $q$ is generic implies that $X_\alpha$ cannot have any zero. Otherwise, it would have a circle of zero which contradicts its covariant holomorphicity. Therefore, $X_\alpha\neq 0$ throughout the cylinder so the phase of $X_\alpha$ (in a global trivialisation of the gauge bundle over $HS^2$) has the same winding numbers at the two boundary circle. Therefore, the relative winding number of the two trivializations is zero. The Taubes equation for $h = 2 \log(|X_\alpha|)$ can be written as
\begin{eqnarray}
          \nabla^2h + 1 - e^h = 0 \, ,
\end{eqnarray}
and has a unique solution by the argument in \cite{ZhaoVortex} and therefore $h = 0$ throughout the cylinder and field strength $F_{z\bar{z}} = 0$. Hence $X_\alpha$ is a pure gauge.

The one-loop determinant for the chiral $X_\beta, \beta\neq \alpha$ can be computed in a similar way as in section 4.3.1 of \cite{ZhaoBLT}. The mode matching 4.20 of \cite{ZhaoBLT} remains valid. Unmatched modes are antiholomorphic modes of the one-form fermion in the matter chiral. On $HS^2$ they have the form $\bar{z}, \bar{z}d\bar{z}, \bar{z}^2d\bar{z}, \bar{z}^3d\bar{z},... $ (eqn 4.24 of \cite{ZhaoBLT}) due to the regularity constraint at the tip $z = 0$ \footnote{The coordinate chart is chosen so that the $U(1)$ rotation acts as $z\to qz$}. The unmatched modes on the cylinder are $\bar{z}^nd\bar{z}, n\in\Z$ as there is no longer any regularity constraint. The $\bar{z}^nd\bar{z}$ mode contributes $\sqrt{q^n a_\beta/a_\alpha} - \sqrt{q^{-n}a_\alpha/a_\beta}$ by eqn 4.25 of \cite{ZhaoBLT}. And we obtain the contribution for the $X_\beta$ chiral: $\prod_{n\in \Z}\sqrt{q^n a_\beta/a_\alpha} - \sqrt{q^{-n}a_\alpha/a_\beta} $.

 The one-loop determinant of $X_\alpha$ is found to cancel the one-loop determinant of the vector multiplet. It may seems that the one-form fermion in $X_\alpha$ has a zero mode $d\bar{z}$ which contributes $\sqrt{a_\alpha/a_\alpha} - \sqrt{a_\alpha/a_\alpha} = 0$. However, the Lagrangian contains cubic couplings between $X_\alpha$, the matter fermions and the gaugino (last four term of equation A.8 of \cite{ZhaoBLT}, where the $\psi$ are matter fermions and $\lambda$ are gauginos). Together with the kinetic term for $\psi$ and $\lambda$ the cubic coupling turns two massless Dirac fermions into two massive Dirac fermions \footnote{In the 4d language, it creates a mass terms for two Weyl fermions to form one massive Dirac fermion.}.
 Hence there is no divergence associated with zero modes.

\subsection{Factorization of the sphere partition functions}
In this section, we show that the $S^2 \times S^1$ partition function can be factorized into hemisphere partition functions with gluing factors determined by the interval partition functions. This is illustrated in fig.~\ref{fig: sphere}.

The twisted $S^2 \times S^1$ partition function was first computed in \cite{Benini3d}.  The result with a Wilson loop $W_i$ inserted at the north pole and a loop $W_j$ at the south pole is
\begin{eqnarray}
	S_{ij}(\Lambda, q, a) = \sum_{d\geq 0}\Lambda^d  \int \frac{ds}{2\pi i s} \frac{(sq^{d})^i s^j}{\prod_{\alpha = 1}^N \prod_{n = 0}^d \sqrt{sq^n a_\alpha}-\sqrt{s^{-1}q^{-n} a_\alpha^{-1}}} \, ,
\end{eqnarray}
where we have uplifted the $U(1)$ action from $S^2$ to the total space of $O(d)\to S^2$ in a way that its weight is $d$  on the fibre above the north pole and $0$ on the fibre over the south pole. The final result is independent of the choice of the uplift. $s$ is the complexified holonomy $\exp(\int -iA_0 + \sigma)$ along the time circle $S^1_q$.  The angular momenta of sections of $H^0(O(d))$ are $1, q, q^2,..., q^{d}$. The Wilson loop at the north pole therefore contributes $sq^d$ while the Wilson loop at the south pole contributes $s$. The integral picks up poles at $s = q^{-n}a_\alpha^{-1}, n = 0,..., d$ in the degree $d$ sector. The result is \footnote{The (absolute) convergence of $S_{ij}$ is guaranteed for $|\Lambda| < 1, 0 < |q| < 1$ due to the (absolute) convergence of the vortex partition function.} (see \cite{BullimoreBVF} and references therein)
\begin{eqnarray}
         S_{ij}(\Lambda, q, a) = 
    \sum_{\alpha = 1}^N H^{\text{vortex}}_{\alpha i}(\Lambda, q, a)H^{\text{vortex}}_{\alpha j}(\Lambda, q^{-1}, a)\frac{1}{\prod_{\beta \neq \alpha}\sqrt{a_\beta/a_\alpha} -\sqrt{a_\alpha/a_\beta} }
\end{eqnarray}
Using the identity \eqref{eq: int_sqed} which relates the interval partition function and the perturbative parts of the hemisphere partition functions, we find the relation 
\begin{eqnarray}\label{eq: fact_sphere_sqed}
	S_{ij}(\Lambda, q, a)=  \sum_{\alpha = 1}^N H_{\alpha i}(\Lambda, q, a)H_{\alpha j}(\Lambda, q^{-1}, a)I_{\alpha\alpha}(q, a)^{-1}
\end{eqnarray}
This can be rewritten as
\begin{eqnarray}\label{eq: fact_sphere_matrix_form}
    S_{ij}(\Lambda, q, a)=  \sum_{\alpha, \beta = 1}^N H_{\alpha i}(\Lambda, q, a)H_{\beta j}(\Lambda, q^{-1}, a)I^{\alpha\beta}(q, a)
\end{eqnarray}
where $I^{\alpha\beta}I_{\beta\gamma}= \delta^\alpha_\gamma$. Since the interval partition function (as a matrix) $I_{\alpha\beta}$ is diagonal, so is its matrix inverse $I^{\alpha\beta}$. We have $I^{\alpha\alpha} = I_{\alpha\alpha}^{-1}$. This identity holds for all $i,j\in \Z$. In the next section, we will restrict to a basis of linearly independent Wilson loops $i, j = 0,..., N-1$.

\subsection{Factorization of the interval partition functions}
Now we are ready to prove that
\begin{eqnarray}\label{eq: int_fact_sqed}
	I_{\alpha\beta}(q, a) = \sum_{i, j = 0}^{N - 1} H_{\alpha i}(\Lambda, q^{-1}, a)S^{ij}(\Lambda, q , a)H_{\beta j}(\Lambda, q, a)
\end{eqnarray}
by inverting \eqref{eq: fact_sphere_sqed}, where $S^{ij}S_{jk} = \delta^i_k$. Here we have assumed that $i, j = 0,..., N - 1$. It suffices to check that $S_{ij}$ is invertible. When $\Lambda$ is small, the leading order term in $S_{ij}$ is
\begin{eqnarray}
	S_{ij}(q, a, \Lambda) = \int \frac{ds}{2\pi i s} \frac{s^{i + j}}{\prod_{\alpha = 1}^N \sqrt{s a_\alpha}-\sqrt{s^{-1} a_\alpha^{-1}}} + O(\Lambda)
\end{eqnarray}
This equation has a clear interpretation in terms of the equivariant K-theory of the Higgs branch $P^{N - 1}$ with the flavour symmetry torus action $X_1,..., X_N\mapsto a_1X_1, ..., a_NX_N$ where $X_i$ are homogeneous coordinates. Indeed, if we identify $s $ with $O(1) \to P^{N- 1}$ where the equivariant structure is such that the fibre above the torus fixed point $X_\alpha = 1, X_i = 0, \forall i \neq \alpha$ has weight $a_\alpha^{-1}$, then the $S^2$ partition function equals the symmetrized K-theoretic inner product of $O(i)$ and $O(j)$.
\begin{eqnarray}
	\chi(O(i) \times O(j) \times \sqrt{KP^{N - 1}})
\end{eqnarray}
This inner product is non-degenerate on the localized equivariant K-theory which has an orthogonal basis consisting of skyscraper sheaves  at the fixed points of the torus action. A more algebraic proof goes as follows:
\begin{eqnarray}
    S_{ij} = \sum_{\alpha = 1}^N \frac{a_\alpha^{-i - j}}{\prod_{\beta \neq \alpha}\sqrt{a_\beta/a_\alpha} - \sqrt{a_\alpha/a_\beta}}
\end{eqnarray}
Now we identify the Wilson loop $W_i$ by the vector $[a_1^{-i}, a_2^{-i},...,a_N^{-i}]\in \C^N$. Then we can write $S_{ij}$ as an inner product $(W_i, W_j)_{K(P^{N - 1})}$ where the $K$-theoretic inner product on $\C^N$ is:
\begin{eqnarray}
    ([c_1,...,c_N], [d_1,..., d_N])_{K(P^{N - 1})} = \sum_{\alpha = 1}^N \frac{c_\alpha d_\alpha}{\prod_{\beta \neq \alpha}\sqrt{a_\beta/a_\alpha} - \sqrt{a_\alpha/a_\beta}}
\end{eqnarray}
The inner product is nondegenerate since its matrix is diagonal and does not contain any zero diagonal entry. Hence, it suffices to prove that the Wilson loops $W_i$ (identified with vectors in $\C^N$ above) form a basis of $\C^N$ as $S_{ij}$ is the Gram matrix. This follows because the matrix formed by these vectors give a Vandermonde determinant which is non-zero.

Since the $N\times N$ matrix $S_{ij}$ for $i,j=0,..., N - 1$ is invertible, all matrices on the right hand side of \eqref{eq: fact_sphere_matrix_form} are invertible. So we can invert the two $H$ matrices and move to the left hand side. This completes the proof of \eqref{eq: int_fact_sqed}.

One can give a more direct argument that our choice of Wilson loops $1, s,..., s^{N - 1}$ (times the correction factor in \eqref{eq: wilson_loop}) form a complete basis of all Wilson loops by using the $S^2$ partition function. If $O_N, O_S$ denote the Wilson loops at the north and south poles respectively, their inner product is
\begin{eqnarray}
         \langle O_N O_S \rangle = \sum_{d\geq 0}\Lambda^d  \int \frac{ds}{2\pi i s} \frac{O_N(sq^{d}) O_S(s)}{\prod_{\alpha = 1}^N \prod_{n = 0}^d \sqrt{sq^n a_\alpha}-\sqrt{s^{-1}q^{-n} a_\alpha^{-1}}} \, ,
\end{eqnarray} 
We have the identity
\begin{eqnarray}
       \langle O_N(s) O_S(s) \prod_{\alpha = 1}^N \left(\sqrt{sa_\alpha} - \sqrt{s^{-1}a_\alpha^{-1}}\right) \rangle = \Lambda \langle O_N(s) O_S(sq^{-1}) \rangle
\end{eqnarray}
This holds for any $O_N(s)$ which means we can identify the following two Wilson loops:
\begin{eqnarray}
          O_S(s) \prod_{\alpha = 1}^N \left(\sqrt{sa_\alpha} - \sqrt{s^{-1}a_\alpha^{-1}}\right) \sim \Lambda O_S(sq^{-1})
\end{eqnarray}
for any Laurent polynomial $O_S(s)$. Since $N$ is even by our assumption, the product consists only of integer powers of $s$. The space of Laurent polynomial in $s$ quotient by this relation is a $N$-dimensional vector space spanned by $1, s,..., s^{N - 1}$. For example, setting $O_S(s) = s^{N / 2}$ allows us to write $s^N$ as a linear combination of $1,s,...,s^{N - 1}$. A similar relation holds for $O_N(s)$.

\section{$U(k)_{-k-N}$ in the presence of boundary matter}\label{sec: cs_ym}
In this section we factorize the interval partition functions of 3d $\mathcal{N} =2$ $U(k)$ Chern-Simons-Yang-Mills theory with boundary matter. We set the Chern-Simons level at $-k-N$, where $N$ is a nonnegative integer. This would force us to add boundary matter  to cancel the boundary anomaly. It is not known how to couple the topologically twisted theory on $HS^2 \times S^1$ to boundary matter while preserving supersymmetry. Therefore, we need to use the rigid supersymmetric hemisphere partition function computed in \cite{DavideDualBC}. We also need to use the interval partition function on $T^2 \times [0, 1]$ computed by Yoshida \cite{YoshidaInterval}, without an $\Omega$-background. The torus $T^2 \times [0, 1]$ carries the product of the flat metric on $\C/(\Z \oplus \Z\tau) \times [0, 1]$.

It is not clear a priori why \eqref{eq: fact_int} should hold in this case. Indeed, the interval partition function does not have any $\Omega$ background while the hemisphere partition function does. We also note that the bulk fermions are periodic along both cycles of the interval geometry $T^2 \times [0, 1]$ but antiperiodic along one of the cycles of the boundary torus of $HS^2 \times [0,1]$. Despite these differences, we shall find that \eqref{eq: fact_int}
still holds with suitable gluing factors. It is not clear whether the gluing factors are related to the $S^2 \times S^1$ partition functions at the moment. We now describe the field content of the theory:

The field content on $T^2 \times [0, 1]$ is:
\begin{enumerate}
	\item[(a)] a 3d $\mathcal{N} = 2$ $U(k)$ vector multiplet $(A_\mu, \sigma)$ with $SU(k)$ Chern-Simons level $-k - N$ and $U(1)$ Chern-Simons level $-N$. The embedding $u(1)\to u(k)$ is given by $1\to I_k/\sqrt{k}$. \cite{DavideDualBC}.
	\item[(b)] $N$ 2d $(0, 2)$ fundamental Fermi multiplets in the NS sector\footnote{This means all the fields in the Fermi multiplets have twisted periodicities along the two independent cycles of the torus. One can also consider the Fermi multiplets in any of the four spin structures of the torus. The auxiliary field needs to have the same periodicity as the fermions to preserve supersymmetry. The partition functions
		      of the different spin structures are related to each other via $z\to z^{-1}$ and $z\to zq^{\pm 1/2}$. For example, $(z;q)_\infty(qz^{-1};q)_\infty$ is related to $(q^{1/2}z;q)_\infty(q^{1/2}z^{-1};q)_\infty$ via $z\to zq^{1/2}$. The Fermi multiplets on the two boundary tori can have different spin structures. We picked the most convenient spin structure to do the factorization.} with $SU(N)$ flavour fugacities $x_i, i = 1,..., N$ on the first boundary torus subject to the constraint $\prod_i x_i = 1$.
	\item[(c)] $N$ 2d $(0, 2)$ fundamental Fermi multiplets in the NS sector with $SU(N)$ flavour fugacities $y_i, i = 1,..., N$ on the second boundary torus subject to the constraint $\prod_i y_i = 1$.
\end{enumerate}
Together we have a product $SU(N) \times SU(N)$ of flavour symmetries rotating the boundary fermions \footnote{We also have a $U(1)$ symmetry rotating the boundary fermions on the two boundary tori in the opposite way. It is possible to assign a fugacity to this symmetry and include it in the factors $f_{\mu \lambda}(q)$. It is also possible to gauge it by adding an additional 3d $\mathcal{N} = 2$ $U(1)$ vector multiplet in the bulk with a suitable Chern-Simons level. In that case, the factorization of the partition function would take a similar form as \eqref{eq:diagonal_fact}.}. The boundary gauge anomaly is cancelled independently on each boundary \cite{DavideDualBC}. The $SU(k)$ Chern-Simons term contributes a factor $-k - N$. The Yang-Mills term contributes $k$ and the $N$ Fermi multiplets contribute $N$. A similar analysis shows that the $U(1)$ gauge anomaly also cancels. When $N = 0$, there is no Fermi multiplet on the boundary and we are left with purely bulk degrees of freedom. We impose the following boundary conditions for the gauge field at both boundary tori:
\begin{eqnarray}\label{eq: bc_cs_ym}
	\partial_0 A_{\bar{z}} = A_0 = 0 \, ,
\end{eqnarray}
which admit a unique $(0,2)$ supersymmetric completion.

We can also place the $U(k)$ vector multiplet on the hemisphere $HS^2\times S^1$ with the same Chern-Simons level and $N$ fundamental boundary Fermi multiplets with flavour fugacities $x = (x_1,..., x_N)$ \cite{DavideDualBC}. We can also switch on a Wilson line in the representation $\lambda$ of $U(k)$. The hemisphere partition function is written as $H_{\lambda}^{N, k}(q, x)$, where $x = (x_1,..., x_N)$. We will prove, using the conformal branching rule $\widehat{U(k)}_N\oplus \widehat{SU(N)}_k\to \widehat{U(kN)}_1$, that this equals the character of $\widehat{SU(N)}_k$ with a different highest weight $\tilde{\lambda}$ up to an overall factor of $q$. The $\tilde{\lambda}(\lambda)$ relation will be given in the main text and follows from the level-rank duality of Chern-Simons theories.

The factorization of the interval partition function into affine characters can be proven using two different methods. The first method relies on a set of difference equations satisfied by the affine characters which form a basis of the solution space. The interval partition functions satisfy these difference equations separately in the two set of fugacities $x, y$ and therefore can be written as a sum of products of affine characters. The second method is more explicit and again uses the branching rule $\widehat{U(k)}_N\oplus \widehat{SU(N)}_k\to \widehat{U(kN)}_1$.

\subsection{Interval Partition Functions}
The interval partition function for a $U(k)$ gauge multiplet coupled to $N$ fundamental Fermi multiplets on each boundary torus described above is
\begin{eqnarray}\label{eq:CSYM_int_pf}
	I^{N, k}(q, x, y) = \frac{1}{k!}\oint_{|s_i| = 1}
	\left(\prod_{i = 1}^k \frac{ds_i}{2\pi i s_i}\right)
	I_{\text{1-loop}}^{k}(q, s)\prod_{i=1}^k\prod_{a=1}^N Z_{\text{ferm}}(q, s_ix_a)Z_{\text{ferm}}(q, s_iy_a^{-1}) \, , \nonumber \\
\end{eqnarray}
where
\begin{eqnarray}
	\begin{aligned}
		I_{\text{1-loop}}^{k}(q, s)  & = \prod_{i\neq j} \left(1 - \frac{s_i}{s_j}\right)\prod_{i, j = 1}^k (q s_i / s_j ; q)_\infty^2 \\
		Z_{\text{ferm}}(q, z) & = (-q^{1/2}z;q)_\infty(-q^{1/2}z^{-1};q)_\infty \, ,
	\end{aligned}
\end{eqnarray}
and
\begin{eqnarray}
	s = (s_1, s_2,..., s_k)\quad x = (x_1, x_2,..., x_N) \quad y = (y_1, y_2, ..., y_N) \quad z\in \C^*
\end{eqnarray}
$I_{\text{1-loop}}^{k}$ is the one-loop determinant of the 3d $U(k)$ gauge multiplet on the interval geometry. $Z_{\text{ferm}}$ is the NS sector partition function of a free complex fermion of charge $z$. The contour is a direct product of the unit circles. The only poles of $s_i$ are at the origin so we can deform the contours as long as they surround the unit circle. We also enforce the constraints
\begin{eqnarray}
	\prod_{i} x_i = \prod_{i}y_i = 1
\end{eqnarray}
as they correspond to demanding having $SU(N)$ fugacities. A direct computation of the contour integral is difficult as the NS partition functions $Z_{\text{ferm}}$ have essential singularities at the origin $s_i = 0$. Therefore, we will use a resummation of the integrand following \cite{DavideDualBC}.

The Chern-Simons level-rank duality $U(k)_{-k - N} \leftrightarrow SU(N)_{k + N}$ suggests that the partition function \eqref{eq:CSYM_int_pf} would factorize into a sum of products of $\widehat{SU(N)}_k$ characters. We will show that this is indeed the case, with the presence of additional normalization factors. In the next section, we will prove that affine characters can be written as hemisphere partition functions. Therefore, the factorization can be regarded as a factorization into hemisphere partition functions.

\subsection{Hemisphere Partition Functions}
The hemisphere partition function of a $U(k)$ vector multiplet coupled to $N$ fundamental boundary Fermi multiplets with $SU(N)$ flavour fugacities $x_i$ and a Wilson line in the representation $\lambda$ of $U(k)$ in the bulk is given by:
\begin{eqnarray}
	H_\lambda^{N, k}(q, x_1,..., x_N) = \frac{1}{k!}\oint_{|s_i| = 1}
	\left(\prod_{i = 1}^k \frac{ds_i}{2\pi i s_i}\right)
	H_{\text{1-loop}}^{k}(s)\prod_{i = 1}^k \prod_{a = 1}^N Z_{\text{ferm}}(q, s_ix_a) \chi_\lambda^{U(k)}(s)
\end{eqnarray}
where
\begin{eqnarray}
	\begin{aligned}
		H_{\text{1-loop}}^{k}(s) = \prod_{i\neq j} \left(1 - \frac{s_i}{s_j}\right)
		\prod_{i, j = 1}^k (q s_i / s_j ; q)_\infty \\
		Z_{\text{ferm}}(q, z) = (-q^{1/2}z;q)_\infty(-q^{1/2}z^{-1};q)_\infty \, .
	\end{aligned}
\end{eqnarray}
In these formulae we have used $s$ to denote $(s_1, s_2,..., s_k)$, the $U(k)$ fugacities. The contour is again the product of unit circles. $H_{\text{1-loop}}^{k}(s)$ is the one-loop determinant of the vector multiplet on the hemisphere. $\chi_\lambda^{U(k)}(s)$ is the character of the representation $\lambda$ of $U(k)$ due to the Wilson line.

The crucial observation which allows us to factorize the interval partition function into hemisphere partition functions is that the hemisphere partition functions compute $\widehat{SU(N)}_k$ characters \cite{DavideDualBC}. This was originally conjectured in \cite{DavideDualBC} as a result of the level-rank duality of Chern-Simons theories. While we expect the proof to be known \footnote{We thank D. Gaiotto for discussions on this proof.}, it has not been explicitly written down in the literature, and we do so in the present work \footnote{The method used in our proof can also be extended to construct integral representations for coset characters of $g/h$ where $g,h$ are simple lie algebras. In these cases, one needs to replace the free fermion partition function by the corresponding character of $g$.}. We would like to prove that the affine character of $\widehat{SU(N)}_k$ with the highest weight $\lambda$ is (notice the shift $s\to s^{-1}$ in $\chi_{\lambda^T}^{U(k)}$)
\begin{eqnarray}\label{eq:wzw_int}
	\begin{aligned}
		\chi_\lambda^{\widehat{SU(N)}_k}(q, x_1,..., x_N) = q^{-|\lambda|/2}\frac{1}{k!}\oint_{|s_i| = 1}
		\prod_{i = 1}^k \frac{ds_i}{2\pi i s_i}
		\prod_{i\neq j} \left(1 - \frac{s_i}{s_j}\right)
		\prod_{i, j = 1}^k (q s_i / s_j ; q)_\infty \\
		\prod_{i = 1}^k \prod_{a = 1}^N (-q^{1/2}s_i x_a; q)_\infty(-q^{1/2}s_i^{-1} x_a^{-1}; q)_\infty \chi_{\lambda^T}^{U(k)}(s^{-1})
	\end{aligned}
\end{eqnarray}
where $s^{-1} = (s_1^{-1},..., s_k^{-1})$. The contour is a direct product of unit circles. $\lambda$ is a representation of $SU(N)$ labelled by a partition (or equivalently a Young diagram) \footnote{We impose the condition $\lambda_N = 0$. The corresponding Young diagram is sometimes called \textit{reduced}. This condition is needed to avoid the possibility that $\lambda^T$ is not a valid $U(k)$ diagram (having too many rows).}
\begin{eqnarray}
	\lambda = (\lambda_1, ..., \lambda_{N - 1}, \lambda_N) \, , \quad \lambda_1\geq \lambda_2 \geq... \geq \lambda_{N - 1}\geq \lambda_N = 0 \, , \quad \lambda_1\leq k \, ,
\end{eqnarray}
and $|\lambda| = \lambda_1 + ... + \lambda_N$ is the number of boxes in $\lambda$. More information can be found in the appendix \ref{app:su(n)}. $\lambda^T$ is the transpose of the Young diagram $\lambda$ and corresponds to a representation of $U(k)$
\footnote{The process $\lambda \to \lambda^T$ is not the only one that leads to the correct $\widehat{SU(N)}_k$ character. Other choices of $U(k)$ characters would also lead to the same result (but with a different prefactor of a power of $q$.) }. The corresponding $U(k)$ character is $\chi_{\lambda^T}^{U(k)}(s_1,..., s_k)$. We list some values of $\chi_{\lambda^T}^{U(k)}$ for small values of $N$ and $k$ in table \ref{tb:transpose_yt}. Throughout this work, we use the convention that affine characters start at $q^0$.

\begin{table}[ht]
	\caption{Values of $\chi_{\lambda^T}^{U(k)}(s)$ for different representations $\lambda$.}
	\label{tb:transpose_yt}
	\centering
	\begin{tabular}{lll}
		\toprule
		Affine Lie algebra  & $\lambda$         & $\chi_{\lambda^T}^{U(k)}$ \\
		\midrule
		$\widehat{SU(2)}_2$ & $\emptyset$       & $1$                       \\
		$\widehat{SU(2)}_2$ & $\ydiagram{1}$    & $s_1 + s_2$               \\
		$\widehat{SU(2)}_2$ & $\ydiagram{2}$    & $s_1s_2$                  \\
		\midrule
		$\widehat{SU(3)}_2$ & $\emptyset$       & $1$                       \\
		$\widehat{SU(3)}_2$ & $\ydiagram{1}$    & $s_1 + s_2$               \\
		$\widehat{SU(3)}_2$ & $\ydiagram{1, 1}$ & $s_1^2 + s_1s_2 + s_2^2$  \\
		$\widehat{SU(3)}_2$ & $\ydiagram{2, 1}$ & $s_1s_2(s_1 + s_2)$       \\
		$\widehat{SU(3)}_2$ & $\ydiagram{2, 0}$ & $s_1s_2$                  \\
		$\widehat{SU(3)}_2$ & $\ydiagram{2, 2}$ & $s_1^2s_2^2$              \\
		\bottomrule
	\end{tabular}
	\label{tab:RT_examples}
\end{table}
For example, take $N=3$ and $k=2$. The symmetric square of the fundamental of $SU(3)$  is represented by the diagram $\ydiagram{2}$. We transpose the diagram to obtain $\ydiagram{1, 1}$ which is the exterior square of the fundamental of $U(2)$ with character $s_1s_2$. The adjoint of $SU(3)$ is represented by the diagram $\ydiagram{2, 1}$. Its transpose is still $\ydiagram{2, 1}$, which is the fundamental of $U(2)$ times the determinant of $U(2)$. Its character is $s_1s_2(s_1 + s_2)$.

Now we begin the proof of \eqref{eq:wzw_int}.

\textbf{STEP I: }The first step of the proof is to rewrite the free fermion partition functions as a power series in the fugacities $s_i$ and $x_a$. To do this, we use the Jacobi triple product/bosonization formula:
\begin{equation}\label{eq:bosonisation}
	(-q^{1/2} x; q)_{\infty} (-q^{1/2} x^{-1}; q)_{\infty} =
	(q; q)_{\infty}^{-1}\sum_{n \in \mathbb{Z}} q^{\frac{1}{2} n^2} x^n \, ,
\end{equation}
to write
\begin{align*}
	  & \prod_{i = 1}^k \prod_{a = 1}^N Z_{\text{ferm}}(q, s_ix_a)                                                                      \\
	= & \prod_{i = 1}^k \prod_{a = 1}^N (-q^{1/2}s_i x_a; q)_\infty(-q^{1/2}s_i^{-1} x_a^{-1}; q)_\infty                                \\
	= & (q;q)_{\infty}^{-Nk}\sum_{n_{ia}\in \Z}q^{\frac{1}{2}\sum_{i, a} n_{ia}^2}\prod_{i = 1}^k \prod_{a = 1}^N(s_ix_a)^{n_{ia}} \, ,
\end{align*}
where each $n_{ia}, 1\leq i\leq k, 1\leq a \leq N$ sums over $\Z$.

Now we insert this expression into the integral \eqref{eq:wzw_int}. We note that only the sector $\sum_{i = 1}^k\sum_{a = 1}^N n_{ia} = |\lambda|$ contributes, where $|\lambda| = |\lambda^T|$ is the number of boxes in $\lambda$. This is due to the symmetry $s_i\mapsto C s_i, \forall i, C \in U(1)$ of the integration measure. The character $\chi_{\lambda^T}^{U(k)}(s^{-1})$ transforms as $C^{-|\lambda|}$. $\prod_{i, a}(s_i x_a)^{n_{ia}}$ transform as $C^{\sum n_{ia}}$. The two must cancel and we arrive at the condition $\sum_{i = 1}^k\sum_{a = 1}^N n_{ia} = |\lambda|$.

\textbf{STEP II: }The next step of the proof decomposes the free fermion characters into characters of $\widehat{U(Nk)}_1$. To do this, we use the free fermion realization of the $\widehat{U(Nk)}_1$ current algebra \cite{DiFrancesco} which states that the sum over $\sum n_{ia} = |\lambda|$ equals the affine character of $\widehat{U(Nk)}_1$ whose zeroth-grade representation is the $|\lambda|$th exterior power of the fundamental:
\begin{eqnarray}\label{eq:free_fermion_real}
	(q;q)_{\infty}^{-Nk}\sum_{\sum n_{ia} = |\lambda|}q^{\frac{1}{2}\sum_{i, a} n_{ia}^2}\prod_{i = 1}^k \prod_{a = 1}^N(s_ix_a)^{n_{ia}} = q^{\frac{1}{2}|\lambda|}\chi_{\wedge^{|\lambda|}\ydiagram{1}}^{\widehat{U(NK)}_1}(q, s_ix_a)
\end{eqnarray}
where the $U(Nk)$ fugacities have been replaced by the products of the $U(k)$ and $SU(N)$ fugacities $s_ix_a, i = 1,..., k, a = 1,..., N$ and $\wedge^{|\lambda|}\ydiagram{1}$ denotes the $|\lambda|$th exterior power of the fundamental of $U(Nk)$.

\textbf{STEP III: } The third step decomposes $\widehat{U(Nk)}_1$ characters in terms of $\widehat{U(k)}_N$ and $\widehat{SU(N)}_k$ characters. To do this, we notice that the right hand side of \eqref{eq:free_fermion_real} naturally leads to the embedding $U(k)\times SU(N)\to U(Nk)$ where the fundamental of $U(Nk)$ becomes the tensor product of the fundamentals of $U(k)$ and $SU(N)$. We now use the affine version of the embedding $\widehat{U(k)}_N\oplus \widehat{SU(N)}_k\to \widehat{U(Nk)}_1$ (which is known to be conformal) to write the $\widehat{U(Nk)}_1$ characters in terms of $\widehat{SU(N)}_k$ and $\widehat{U(k)}_N$ characters. The $\widehat{U(k)}_N$ characters will be integrated out in the integral and we will be left with the $\widehat{SU(N)}_k$ character only.

The conformal branching rule of the embedding $\widehat{U(k)}_N\oplus \widehat{SU(N)}_k\to \widehat{U(Nk)}_1$ is known to be:
\begin{eqnarray}\label{eq:branching}
	\chi_{\wedge^{|\lambda|}\ydiagram{1}}^{\widehat{U(Nk)}_1}(q, s_ix_a) =  \chi_\lambda^{\widehat{SU(N)}_k}(q, x_1,..., x_N) \chi_{\lambda^T}^{\widehat{U(k)}_N}(q, s_1,..., s_k) + ...
\end{eqnarray}
where the remaining terms are strictly different from the leading order term $\chi_{\lambda}(x)\chi_{\lambda^T}(s)$ in the following sense: if $\mu \otimes \kappa$ is a term in the $...$, where $\mu$ and $\kappa$ are integrable representations of $SU(N)$ at level $k$ and $U(k)$ at level $N$ respectively, then $\lambda \neq \mu$ AND $\lambda^T \neq \kappa$ (theorem 1 and lemma 5 of \cite{Altschuler:1989nm}) \footnote{In fact, a stronger result is true. The $SU(k)$ representations of $\kappa$ and $\lambda^T$ are distinct.}. We will soon show that this property implies that only the leading order term $\chi_\lambda(x)\chi_{\lambda^T}(s)$ contributes to the integral \eqref{eq:wzw_int}. Here is an example that illustrates this behaviour (with $\lambda = \ydiagram{1}$):
\begin{equation}
	\begin{aligned}
		  & \chi^{\widehat{U(6)}_1}_{\ydiagram{1}}(q, s_1x_1, s_1x_2, s_1x_3, s_2x_1, s_2x_2, s_2x_3)                                                       \\
		= & \chi^{\widehat{SU(3)}_2}_{\ydiagram{1}}(q, x_1, x_2, x_3) \chi^{\widehat{U(2)}_3}_{\ydiagram{1}}(q, s_1, s_2)                                   \\
		  & + q\chi^{\widehat{SU(3)}_2}_{\ydiagram{2, 2}}(q, x_1, x_2, x_3)\chi^{\widehat{U(2)}_3}_{\ydiagram{3} \otimes \ydiagram{1, 1}^{-1}}(q, s_1, s_2)
	\end{aligned}
\end{equation}
We can see that the two $SU(3)$ representations ($\ydiagram{1}$ and $\ydiagram{2, 2}$) are distinct. The $U(2)$ representations ($\ydiagram{1}$ and $\ydiagram{3} \otimes \ydiagram{1, 1}^{-1}$) are also distinct, where $\ydiagram{3} \otimes \ydiagram{1, 1}^{-1}$ denotes the tensor product of the symmetric cube of the fundamental and the antideterminant representation. The extra factor of $q$ in the last line is needed as all the affine characters start at $q^0$.

\textbf{STEP IV:} The final step of the proof uses the following form of the $\widehat{U(k)}_N$ character proved in appendix \ref{app:inner_product}:
\begin{eqnarray}\label{eq:ch_expansion}
	\chi_{\lambda^T}^{\widehat{U(k)}_N}(q, s_1,..., s_k) = \frac{\chi_{\lambda^T}^{U(k)}(s_1,..., s_k) + ...}{\prod_{i, j = 1}^k\left(qs_i/s_j;q\right)_\infty}
\end{eqnarray}
The numerator is a sum of $U(k)$ characters weighted by powers of $q$. They are of the form $q^{\text{some power}}\chi_\mu^{U(k)}(s_1,..., s_k)$. The $...$ here consists of $U(k)$ characters which are different from $\lambda^T$.\footnote{An alternative method for proving this proposition is to use the formula $L_0(\mu) = (\mu, \mu + 2\rho)/(2(k + N))$ for the conformal weight of the WZW primaries transforming in the representation $\mu$ of $SU(N)$. $\rho$ is the Weyl vector (appendix \ref{app:su(n)}). The $L_0$ of the null states in a Verma module are strictly bigger than the $L_0$ of the highest weight states. Therefore, the null states must transform in a different representation than that of the highest weight states in a Verma module. The numerator of \eqref{eq:ch_expansion} gives a resolution of the irreducible module using Verma modules (a long exact sequence whose last two terms are the irreducible module and 0). We perform induction on the Verma modules in the resolution to show that their highest weights transform in different representations of $\lambda$.} We will soon see that these terms do not contribute to the integral.
Now the Pochhammer in \eqref{eq:wzw_int} clears the denominator $\left(qs_i/s_j;q\right)_\infty$ and the $q^{|\lambda|/2}$ in \eqref{eq:free_fermion_real} cancels the prefactor in \eqref{eq:wzw_int}. Therefore, \eqref{eq:wzw_int} reduces to
\begin{align*}
	  & \frac{1}{k!}\oint_{|s_i| = 1} \prod_{i = 1}^k \frac{ds_i}{2\pi i s_i} \prod_{i\neq j}\left(1 - \frac{s_i}{s_j}\right)\chi_\lambda^{\widehat{SU(N)}_k}(q, x_1,..., x_N) \chi_{\lambda^T}^{U(k)}(s)\chi_{\lambda^T}^{U(k)}(s^{-1}) \\
	= & \chi_\lambda^{\widehat{SU(N)}_k}(q, x_1,..., x_N)
\end{align*}
where we have ignored all the subleading terms in the branching rule \eqref{eq:branching} and the character expansion \eqref{eq:ch_expansion}. To justify the latter, we use the fact that the subleading term in the character expansion \eqref{eq:ch_expansion} consists of $U(k)$ characters of different representations $\mu \neq \lambda^T$ (proved in appendix \ref{app:inner_product}) and the orthogonality of the $U(k)$ characters with respect to the $U(k)$ Haar measure. To justify the former, we write the $\widehat{U(k)}_N$ affine characters in the subleading terms of the branching rule \eqref{eq:branching} in the form \eqref{eq:ch_expansion}. A result from appendix \ref{app:inner_product} states that if $\lambda\neq \mu$ are two integrable representations of $\widehat{U(k)}_N$, the $U(k)$ characters appearing in the numerators of the expansions \eqref{eq:ch_expansion} of the two affine characters are all distinct. Therefore, the orthogonality of $U(k)$ characters with respect to the Haar measure can be used to show that the subleading terms in the branching rule \eqref{eq:branching} do not contribute to the integral \eqref{eq:wzw_int}. The proof is now complete. From now on, we will not distinguish hemisphere partition functions of Chern-Simons-Yang-Mills theories from affine characters of $\widehat{SU(N)}_k$. We will factorize the interval partition function \eqref{eq:CSYM_int_pf} into affine characters.

\subsection{Difference Equations of Affine Characters}
The integral representation of the affine character in the previous section allows us to derive a set of difference equations \footnote{See \cite{Axelrod} for related discussions and the relation to Chern-Simons wavefunctions on a torus. We will comment on the latter in a footnote.} satisfied by affine characters which will be useful in the factorization of the interval partition function.

To derive the difference equations, we again use the notation of the free fermion partition function in the NS sector:
\begin{eqnarray}
	Z_{\text{ferm}}(q, s_ix_a) = (-q^{1/2}s_i x_a;q)_\infty(-q^{1/2}s_i^{-1} x_a^{-1};q)_\infty
\end{eqnarray}
which satisfies
\begin{equation}\label{eq:ferm_par_diff}
	\begin{aligned}
		Z_{\text{ferm}}(q, s_i(qx_a)) =  & q^{-1/2}s_i^{-1}x_a^{-1}Z_{\text{ferm}}(q, s_i x_a) \\
		Z_{\text{ferm}}(q, s_i(x_a/q)) = & q^{-1/2}s_i x_a Z_{\text{ferm}}(q, s_ix_a)
	\end{aligned}
\end{equation}
No product/summation is performed on the indices $i$ and $a$. These relations and \eqref{eq:wzw_int} imply the following difference equations (for any $\lambda$ and $1\leq i < j \leq N$):
\begin{eqnarray}\label{eq:diff_eq}
	\chi^{\widehat{SU(N)}_k}_{\lambda}(q, x_1,..., qx_i,..., q^{-1}x_j,..., x_N) = q^{-k}x_j^k x_i^{-k}\chi^{\widehat{SU(N)}_k}_{\lambda}(q, x_1, x_2, x_3, x_4,..., x_N)
\end{eqnarray}
We have multiplied $x_i$ by $q$ and $x_j$ by $q^{-1}$ only in the left hand side. The constraint $\prod_{i = 1}^N x_i = 1$ is preserved under this operation. There are $N(N - 1)/2$ difference equations labelled by $(i, j)$ where $1\leq i < j \leq N$. Since $\chi_\lambda^{\widehat{SU(N)}_k}$ is completely symmetric in the $x_i$, there is only one independent difference equation. Every other difference equation can be derived by permuting the $x_i$.

The important point is that affine characters form a basis of the solutions to the difference equations under the additional assumption that a solution needs to be completely symmetric in the $x_i$ \footnote{The proof of this fact follows from an isomorphism between the divisor $\prod_{i = 1}^N x_i = 1\subset \text{Sym}^N (\C^*/q^\Z)$ and the projective space $P^{N - 1}$, where $x_i$ are the coordinates on the torus $\C^*/q^\Z$. $\text{Sym}^N (\C^*/q^\Z)$ denotes the $N$th symmetric product of the torus $\C^*/q^\Z$. The affine characters are invariant under permutations of the $x_i$ and so descend to functions (holomorphic sections to be precise) on the symmetric product. The difference equations imply that the affine characters transform as holomorphic sections of a degree $k$ line bundle over this divisor. This is consistent with the interpretations of affine characters as wavefunctions of $SU(N)$ Chern-Simons theory on a torus. In this picture, $x_i$ parametrize complexified holonomies of flat $SU(N)$ connection on the torus and the map $x_i\to x_i q$ corresponds to large gauge transformations of the Chern-Simons theory. We thank Xuanchun Lu and Zhipu Zhao for explaining the proof of this fact to us. }. The interval partition function will be shown to satisfy the same set of difference equations in $x$ and $y$ independently. Therefore, it can be written as a sum of products of affine characters with additional normalization factors. An alternative derivation of the difference equations will be presented in appendix \ref{app:diff_eq}.

\subsection{Factorization}
Now we proceed to factorize the interval partition function \eqref{eq:CSYM_int_pf}:
\begin{eqnarray}
	\begin{aligned}
		I^{N, k}(q, x, y) = \frac{1}{k!}\oint_{|s_i| = 1}
		\prod_{i = 1}^k \frac{ds_i}{2\pi i s_i}
		\prod_{i\neq j} \left(1 - \frac{s_i}{s_j}\right)
		\prod_{i, j = 1}^k (q s_i / s_j; q)_\infty^2                                                     \\
		\prod_{i = 1}^k \prod_{a = 1}^N (-q^{1/2}s_i x_a; q)_\infty(-q^{1/2}s_i^{-1} x_a^{-1}; q)_\infty \\
		\prod_{i = 1}^k \prod_{a = 1}^N (-q^{1/2}s_i y_a^{-1}; q)_\infty(-q^{1/2}s_i^{-1} y_a; q)_\infty
	\end{aligned}
\end{eqnarray}
Again we have used the abbreviation $x = (x_1,..., x_N), y  = (y_1,..., y_N)$. The integration contour is a product of unit circles, one for each $s_i$.

Before dealing with the general case $N\geq 0$, we briefly comment on the case $N = 0$ (no boundary Fermi multiplets). The partition function can be exactly computed (end of appendix \ref{app:inner_product}) \footnote{The interval chiral algebra of this theory is a gauged version of the $\beta,\gamma$ system (coming from the bulk vector multiplet) and $2Nk$ complex fermions. The interval partition function counts gauge invariant operators. It would be good to know if an analogous BRST operator in \cite{Beem} can be used to define a BRST reduction of the chiral algebra (see also \cite{Coman2023}). It would also be good to clarify the relationship of this chiral algebra with any coset construction. It is unclear to the authors whether $f_{\mu\lambda}(q)$ are the branching functions of certain cosets.}
\begin{align*}
	I^{N = 0, k}(q) & = \frac{1}{k!}\oint_{|s_i| = 1}
	\prod_{i = 1}^k \frac{ds_i}{2 \pi i s_i}
	\prod_{i\neq j} \left(1 - \frac{s_i}{s_j}\right)
	\prod_{i, j = 1}^k (q s_i / s_j; q)_\infty^2                                                       \\
	                             & = (q;q)^2_\infty\sum_{r \in \Lambda_R(U(k))} q^{k(r,r)+ 2(\rho, r)} \\
	                             & = (q;q)^2_{\infty}(1 + q^2 + ...)
\end{align*}
where sum is over the $r = (r_1,..., r_k)\in \Z^k$ and $\sum_i r_i = 0$ (the root lattice $\Lambda_R$ of $U(k)$). The Weyl vector is:
\begin{eqnarray}
	\rho = (\frac{k - 1}{2}, \frac{k-3}{2}, \frac{k-5}{2}..., \frac{1 - k}{2})
\end{eqnarray}
and the inner product between vectors is the usual Euclidean inner product. For example, $(r, r) = \sum_{i = 1}^k r_i^2$. The $q^2$ term in the last line is the first nonzero power of $q$ coming from the sum and corresponds to $r = (-1, 0, 0,..., 0, 1)$. The corresponding hemisphere partition function for $N = 0$ (pure Chern-Simons-Yang-Mills without boundary Fermi multiplets) equals $(q;q)_\infty$ \cite{DavideDualBC}. Therefore, $I^{N=0,k}(q)$ is not a product of two copies of the hemisphere partition function. There is an additional normalization factor:
\begin{eqnarray}
	\sum_{r\in \Lambda_R(U(k))} q^{k(r,r) + 2(\rho, r)} \, .
\end{eqnarray}
We will see similar normalization factors for the general case.

Now we move onto the case $N > 0$. We first prove the existence of the factorization of \eqref{eq:CSYM_int_pf} into affine characters using a set of difference equations. The formulas \eqref{eq:ferm_par_diff} imply that $I^{N,k}(q, x, y)$ satisfy the difference equations in $x$ and $y$ separately:
\begin{eqnarray}
	\begin{aligned}
		I^{N, k}(q, qx_1, q^{-1}x_2..., x_N, y_1,..., y_N) = q^{-k}x_2^{k}x_1^{-k}I^{N, k}(q, x_1, x_2,..., x_N, y_1,..., y_N) \\
		I^{N, k}(q, x_1, x_2..., x_N, qy_1, q^{-1}y_2,..., y_N) = q^{-k}y_2^{k}y_1^{-k}I^{N, k}(q, x_1,..., x_N, y_1, y_2, ..., y_N)
	\end{aligned}
\end{eqnarray}
Similar equations hold if we multiply $x_i$ (or $y_i$) by $q$ and $x_j$ (or $y_j$) by $q^{-1}$ provided $i\neq j$.
As the affine characters form a complete basis of solutions to the difference equations, there exists functions $f_{\mu \lambda}^{N, k}(q)$ such that
\begin{eqnarray}
	I^{N, k}(q, x, y) = \sum_{\mu,\lambda} f_{\mu \lambda}^{N, k}(q) \chi_{\mu}^{\widehat{SU(N)}_k}(q, x) \chi_{\lambda}^{\widehat{SU(N)}_k}(q, y)
\end{eqnarray}
where $\mu, \lambda$ sums over integrable representations of $SU(N)$ at level $k$. The existence proof is complete. We now describe an algorithm for computing $f_{\mu\lambda}^{N, k}(q)$. The algorithm shows that $f_{\mu\lambda}(q)$ is nonzero only when $\mu = \lambda$. Hence the factorization is diagonal:
\begin{eqnarray}\label{eq:diagonal_fact}
	I^{N, k}(q, x, y) = \sum_{\mu} f_{\mu \mu}^{N,k}(q) \chi_{\mu}^{\widehat{SU(N)}_k}(q, x) \chi_{\mu}^{\widehat{SU(N)}_k}(q, y)
\end{eqnarray}
We will implement this algorithm when $N$ and $k$ are small.

The algorithm for computing $f_{\mu\lambda}(q)$ consists of the following steps:
\begin{enumerate}
	\item[(a)] Replace the two $Z_{\text{ferm}}$ in \eqref{eq:CSYM_int_pf} by sums of products of $\widehat{SU(N)}_k$ and $\widehat{U(k)}_N$ affine characters using the free fermion realization \eqref{eq:free_fermion_real} and the branching rule \eqref{eq:branching}.
	\item[(b)] The $\widehat{SU(N)}_k$ characters do not participate in the integral of \eqref{eq:CSYM_int_pf}. It suffices to perform the integral \eqref{eq:CSYM_int_pf} for products of $\widehat{U(k)}_N$ characters. These integrals are performed in appendix \ref{app:inner_product}. In particular, only the products of identical $\widehat{U(k)}_N$ characters contribute. It was shown in \cite{Altschuler:1989nm} that different $\widehat{U(k)}_N$ characters are paired with different $\widehat{SU(N)}_k$ characters in \eqref{eq:branching}. Therefore, the factorization of the interval partition function \eqref{eq:diagonal_fact} is diagonal.
\end{enumerate}
The formulas for $f_{\mu\lambda}^{N, k}(q)$ are involved in general as the branching rule $\widehat{U(k)_N}\oplus \widehat{SU(N)_k}\to \widehat{U(kN)}_1$ becomes complicated for large $N$ and $k$. We will focus on special cases where one can derive explicit formulas for $f_{\mu\lambda}(q)$.

First we consider a $U(1)$ gauge theory in the bulk coupled to $N$ fundamental Fermi multiplets on both boundaries ($k = 1$ and $N$ arbitrary). The interval partition function is:
\begin{equation}\label{eq:int_pf_k=1}
	\begin{aligned}
		I^{N, k = 1}(q, x, y) = & (q;q)_\infty^2 \oint_{|s| = 1}  \frac{ds}{2\pi i s}   \prod_{a = 1}^N (-q^{1/2}s x_a; q)_\infty(-q^{1/2}s^{-1} x_a^{-1}; q)_\infty \\
		                                     & \quad\quad\quad\quad\quad\quad \prod_{a = 1}^N (-q^{1/2}s y_a^{-1}; q)_\infty(-q^{1/2}s^{-1} y_a; q)_\infty
		\\
		=                                    & (q;q)_\infty^{2 - 2N}\sum_{\sum n_i = \sum m_i} q^{(\sum n_i^2 + \sum m_i^2)/2}x_1^{n_1}...x_{N}^{n_N} y_1^{m_1}... y_N^{m_N}
	\end{aligned}
\end{equation}
where we have used \eqref{eq:bosonisation} and perform the integral with respect to $s$. The $n_i$ are integers and come from applying  \eqref{eq:bosonisation} to the first product in \eqref{eq:int_pf_k=1}. The $m_i$ are also integers and come from the second product in \eqref{eq:int_pf_k=1}. The integral over $s$ enforces the constraint $\sum_{i = 1}^Nn_i = \sum_{i = 1}^N m_i$ in the sum. Now we express the sum with $\sum n_i = \sum m_i = aN + b$ where $a\in \Z, 0\leq b \leq N - 1$ using $\chi^{\widehat{SU(N)}_1}_{\wedge^b \ydiagram{1}}$, the affine character of $\widehat{SU(N)}_1$ where the highest weight is the $b$th exterior power of the fundamental. We do the $x$-sector first, keeping in mind that all affine characters start at $q^0$ in our convention:
\begin{equation}
	\begin{aligned}
		  & (q;q)_{\infty}^{1 - N}\sum_{\sum n_i = aN + b} q^{(\sum n_i^2) / 2}x_1^{n_1}... x_N^{n_N}                          \\
		= & (q;q)_{\infty}^{1 - N} \left(\sum_{\sum n_i = b} q^{(\sum n_i^2) / 2}x_1^{n_1}... x_N^{n_N}\right) q^{Na^2/2 + ab} \\
		= & q^{b/2} \chi^{\widehat{SU(N)}_1}_{\wedge^b \ydiagram{1}}(q, x) q^{Na^2/2 + ab}
	\end{aligned}
\end{equation}
As a result \footnote{If we use the alternative normalization convention for the affine character which start with $q^{h_\lambda}$ where $h_\lambda = (b - b^2/N)/2$ is the conformal weight of the primary of $\widehat{SU(N)}_1$ transforming in $\wedge^b\ydiagram{1}$, the formulas here admit more elegant forms. The normalization factors become exactly the $U(1)$ lattice sums $\sum_{a\in \Z}q^{N(a + b/N)^2}$.},
\begin{align*}
	I^{N, k = 1}(q, x, y) = & \chi_{\emptyset}^{\widehat{SU(N)}_1}(q, x)\chi_{\emptyset}^{SU(N)_1}(q, y)\sum_{a\in\Z} q^{Na^2}                                                        \\
	                                 & + \chi_{\ydiagram{1}}^{\widehat{SU(N)}_1}(q, x)\chi^{\widehat{SU(N)}_1}_{\ydiagram{1}}(q, y)\sum_{a\in \Z} q^{Na^2 + 2a + 1} + ...                      \\
	=                                & \sum_{b = 0}^{N - 1}\chi_{\wedge^b \ydiagram{1}}^{\widehat{SU(N)}_1}(q, x)\chi_{\wedge^b \ydiagram{1}}^{SU(N)_1}(q, y)\sum_{a\in \Z} q^{Na^2 + 2ab + b}
\end{align*}
from which we can read off
\begin{eqnarray}
	f_{\wedge^b\ydiagram{1}, \wedge^b\ydiagram{1}}^{N, k = 1}(q) = \sum_{a\in \Z} q^{Na^2 + 2ab + b}
\end{eqnarray}
where $\ydiagram{1}$ denotes the fundamental of $SU(N)$ and $\wedge^b\ydiagram{1}$ denotes the $b$th exterior power of the fundamental.

The general case is similar. First we write the free fermion partition functions in terms $\widehat{U(Nk)}_1$ characters.
\begin{align*}
	  & \prod_{i = 1}^k\prod_{a = 1}^N (-q^{1/2}s x_a; q)_\infty(-q^{1/2}s^{-1} x_a^{-1}; q)_\infty \\
	= & \sum_{b = 0}^{Nk - 1}\chi_{\wedge^b \ydiagram{1}}^{\widehat{U(Nk)}_1}(q, sx)
	\times \sum_{a\in \Z}
	q^{Nka^2/2 + ab + b/2}\prod_{i = 1}^k s_i^{aN}
\end{align*}
Since the integrand must be invariant under $s_i\to C s_i$ for any $C\in U(1)$, the $x$-sector with fixed $a, b$ must be matched to the $y$-sector with the same $a, b$.
\begin{eqnarray}\label{eq:int_pf_alt_form}
	\begin{aligned}
		I^{N, k}(q, x, y) = \frac{1}{k!}\oint_{|s_i| = 1}
		\prod_{i = 1}^k \frac{ds_i}{2\pi i s_i}
		\prod_{i\neq j} \left(1 - \frac{s_i}{s_j}\right)
		\prod_{i, j = 1}^k (q s_i / s_j; q)_\infty^2 \\
		\sum_{b = 0}^{Nk - 1} \chi_{\wedge^b\ydiagram{1}}^{\widehat{U(Nk)}_1}(q, sx)
		\chi_{\wedge^b\ydiagram{1}}^{\widehat{U(Nk)}_1}(q, s^{-1}y)\sum_{a\in \Z}
		q^{Nka^2 + 2ab + b}
	\end{aligned}
\end{eqnarray}
Now we apply the branching rule \eqref{eq:branching}. We use the case $k = 2, N = 2$ to illustrate the computations of $f_{\mu\lambda}(q)$. The branching rules are:
\begin{equation}\label{eq:u(4)_branching}
	\begin{aligned}
		\chi_{\emptyset}^{\widehat{U(4)}_1}(q, sx) =          & \chi_{\emptyset}^{\widehat{U(2)}_2}(q, s)\chi_{\emptyset}^{\widehat{SU(2)}_2}(q, x) + q\chi_{\ydiagram{2}\otimes \ydiagram{1,1}^{-1}}^{\widehat{U(2)}_2}(q, s)\chi_{\ydiagram{2}}^{\widehat{SU(2)}_2}(q, x) \\
		\chi_{\ydiagram{1}}^{\widehat{U(4)}_1}(q, sx) =       & \chi_{\ydiagram{1}}^{\widehat{U(2)}_2}(q, s)\chi_{\ydiagram{1}}^{\widehat{SU(2)}_2}(q, x)                                                                                                                   \\
		\chi_{\ydiagram{1, 1}}^{\widehat{U(4)}_1}(q, sx) =    & \chi_{\ydiagram{2}}^{\widehat{U(2)}_2}(q, s)\chi_{\emptyset}^{\widehat{SU(2)}_2}(q, x) + \chi_{\ydiagram{1, 1}}^{\widehat{U(2)}_2}(q, s)\chi_{\ydiagram{2}}^{\widehat{SU(2)}_2}(q, x)                       \\
		\chi_{\ydiagram{1, 1, 1}}^{\widehat{U(4)}_1}(q, sx) = & \chi_{\ydiagram{2, 1}}^{\widehat{U(2)}_2}(q, s)\chi_{\ydiagram{1}}^{\widehat{SU(2)}_2}(q, x)
	\end{aligned}
\end{equation}
where we have used the shorthand $sx = (s_i x_a), 1\leq i\leq k, 1\leq a \leq N$ and $s = (s_1,..., s_k), x= (x_1,...x_N)$. In our case $k = N = 2$. The $U(2)$ representation $\ydiagram{2}\otimes \ydiagram{1,1}^{-1}$ refers to the tensor product of the symmetric square of the fundamental and the antideterminant. Using
\begin{eqnarray}
	\wedge^0 \ydiagram{1} = \emptyset\quad \wedge^1 \ydiagram{1} = \ydiagram{1}\quad \wedge^2 \ydiagram{1} = \ydiagram{1,1} \quad \wedge^3\ydiagram{1} = \ydiagram{1, 1, 1}
\end{eqnarray}
for $U(4)$, we can substitute \eqref{eq:u(4)_branching} into \eqref{eq:int_pf_alt_form}, perform the integrals over $\widehat{U(2)}_2$ characters and obtain:
\begin{align*}
	I^{N = 2, k = 2}(q, x, y) = & \chi_{\emptyset}^{\widehat{SU(2)}_2}(q, x)\chi_{\emptyset}^{\widehat{SU(2)}_2}(q, y)f_{\emptyset,\emptyset}^{N = 2, k = 2}(q)               \\& + \chi_{\ydiagram{1}}^{\widehat{SU(2)}_2}(q, x)\chi_{\ydiagram{1}}^{\widehat{SU(2)}_2}(q, y)f^{N = 2, k = 2}_{\ydiagram{1},\ydiagram{1}}(q) \\
	                                         & + \chi_{\ydiagram{2}}^{\widehat{SU(2)}_2}(q, x)\chi_{\ydiagram{2}}^{\widehat{SU(2)}_2}(q, y)f^{N = 2, k = 2}_{\ydiagram{2},\ydiagram{2}}(q)
\end{align*}
where the factors $f_{\mu\lambda}^{N = 2, k = 2}(q)$ are given by \footnote{An anonymous referee pointed out these formulas may equal products of theta functions. It would be interesting to check if they can be further simplified and provide physical explanations.}:
\begin{align*}
	f_{\emptyset,\emptyset}^{N = 2, k = 2}(q) =        & |\chi_{\emptyset}^{\widehat{U(2)}_2}|^2  \sum_{a\in \Z} q^{4a^2} + |\chi_{\ydiagram{2}}^{\widehat{U(2)}_2}|^2\sum_{a\in \Z} q^{4a^2 + 4a + 2}               \\
	f^{N = 2, k = 2}_{\ydiagram{1},\ydiagram{1}}(q) =  & |\chi_{\ydiagram{1}}^{\widehat{U(2)}_2}|^2 \sum_{a\in \Z} q^{4a^2 + 2a + 1} + |\chi_{\ydiagram{2, 1}}^{\widehat{U(2)}_2}|^2\sum_{a\in \Z} q^{4a^2 + 6a + 3} \\
	f^{N = 2, k = 2}_{\ydiagram{2}, \ydiagram{2}}(q) = & q^2 \sum_{a\in \Z} q^{4a^2} |\chi_{\ydiagram{2}}^{\widehat{U(2)}_2}|^2 +  \sum_{a\in \Z} q^{4a^2 + 4a + 2} |\chi_{\emptyset}^{\widehat{U(2)}_2}|^2
\end{align*}
where we have used the inner product defined in the appendix \eqref{eq:inner_product_affine_ch}.
\begin{eqnarray}\label{eq:ex_inner_prod_affine_ch}
	|\chi_{\lambda = (\lambda_1, \lambda_2)}^{\widehat{U(2)}_2}|^2 = \sum_{r_1 + r_2 = 0, r_1\in \Z}q^{4(r_1^2 + r_2^2) - 2r_1(\lambda_1 + 1/2) - 2r_2(\lambda_2 - 1/2)}
\end{eqnarray}
which follows from setting
\begin{eqnarray}
	\begin{aligned}
		r = (r_1, r_2)\quad \rho = (1/2, -1/2)\quad \lambda = (\lambda_1, \lambda_2)\quad N + k = 4 \\
		(r, r) = r_1^2 + r_2^2\quad (r, \lambda + \rho) = r_1(\lambda_1 + 1/2) + r_2(\lambda_2 - 1/2)
	\end{aligned}
\end{eqnarray}
in \eqref{eq:inner_product_affine_ch} and the fact that
\begin{eqnarray}
	|\chi_{\ydiagram{2}}^{\widehat{U(2)}_2}|^2 = |\chi_{\ydiagram{2} \otimes \ydiagram{1,1}^{-1}}^{\widehat{U(2)}_2}|^2\quad |\chi_{\emptyset}^{\widehat{U(2)}_2}|^2 = |\chi_{ \ydiagram{1,1}}^{\widehat{U(2)}_2}|^2
\end{eqnarray}

\section{Discussion}

In this work, we factorized interval partition functions (transition amplitudes) of 3d $\mathcal{N} = 2$ gauge theories into sums of products of elementary building blocks (``hemisphere'' partition functions). We used supersymmetric QED and Chern-Simons-Yang-Mills theories as examples. We also proved the conjecture of \cite{DavideDualBC} that the hemisphere partition functions of Chern-Simons-Yang-Mills are equal to affine characters. The factorization leads to a natural inner product on the space of affine characters at a given level. We proved that affine characters are orthogonal with respect to this inner product.

It would be interesting to extend our analysis to similar setups across different dimensions. It is also important to include boundary degrees of freedom living on the end-points of the interval which are coupled to the bulk degrees of freedom (see \cite{YoshidaInterval,Betzios:2021fnm}). We initiated this study in the case of CS-YM theory by coupling it to boundary fundamental Fermi multiplets and found that the interval partition function can be factorized into boundary wavefunctions/affine-characters with an additional dependence on the fugacities for the global symmetries of the Fermi multiplets.

Finally it would be very interesting to study examples where either the bulk or the boundary field theories are holographic QFTs. A first step towards this goal is to understand the bulk dual of hemisphere partition functions with Wilson loop insertions for large-N strongly coupled holographic gauge theories as well as the resulting interval partition functions. We expect an analogous physical picture to that in \cite{Beem:2012mb,Nieri:2015yia}, where the holomorphic blocks that the partition functions factorize into, exhibit a gravitational block description found in \cite{Hosseini:2019iad,Hosseini:2021mnn}. Such a relationship is expected to hold in the appropriate semi-classical Cardy (large central charge) limit. This analysis would help to clarify whether it is possible to construct two boundary (Euclidean wormhole) saddles in supergravity, starting from more basic building blocks (like the $\langle B | \alpha \rangle$ sectors) and performing diagonal sums, as conjectured in~\cite{Betzios:2021fnm,Betzios:2023obs}.

\acknowledgments
We thank Davide Gaiotto, Bob Knighton, Vit Sripachakul, Xuanchun Lu, Zhipu Zhao, Daniel Zhang, David Tong, Nick Dorey, Matthias Gaberdiel, Andrei Negut, Andrea Ferrari, David Skinner for useful discussions. We thank Samuel Crew, Cyril Closset and Mathew Bullimore for reading through the draft and providing useful feedback. BZ would like to thank GPT-5.5 for pointing out the correction term in the supersymmetric Wilson loop. This work has been partially supported by STFC consolidated
grant ST/X000664/1. P.B. acknowledges financial support from the European Research Council (grant BHHQG-101040024), funded by the European Union. Views and opinions expressed are those of the authors only and do not necessarily reflect those of the European Union or the European Research Council. Neither the European Union nor the granting authority can be held responsible for them.

\appendix

\section{$U(N)$ versus $SU(N)$}\label{app:su(n)}
In this appendix, we collect our convention for young tableaux and affine characters for $U(N)$ and $SU(N)$. A $U(N)$ weight is given by $\lambda = (\lambda_1,..., \lambda_N)$ where all the $\lambda_i$ are integers. The weight $\lambda$ corresponds to $\sum_{i = 1}^N \lambda_i z_i$ where $z_1,..., z_N$ are the diagonal elements of the Cartan subalgebra (of the Lie algebra) of $U(N)$. An $SU(N)$ weight $\lambda$ is given an integer vector of the same form. The formula $\sum_{i = 1}^N \lambda_i z_i$ still holds. The only difference is that we now impose the constraint $\sum_{i = 0}^N z_i = 0$ as the Lie algebra of $SU(N)$ is traceless.
Therefore, two $SU(N)$ weights are the same if they differ by a multiple of $(1,1,..., 1)$. A weight $\lambda$ is called a highest weight if $\lambda_1\geq \lambda_2... \geq \lambda_N$. Any real vector which satisfies this constraint is said to lie in the fundamental chamber. The highest weights of $U(N)$ and $SU(N)$ can be represented by Young diagrams whenever $\lambda_N \geq 0$ (which can always be achieved for $SU(N)$ by adding a multiple of $(1,1,..., 1)$).
The number of boxes in each row of the corresponding Young diagram is $\lambda_i$. For example,
\begin{eqnarray}
	\lambda = (5,3,2)\to \ydiagram{5,3,2}\quad  \lambda = (4, 1)\to \ydiagram{4, 1}
\end{eqnarray}
In the $U(N)$ case, different Young diagrams label different highest weights and correspond to different representations. In the $SU(N)$ case, two Young diagrams label the same representation if they differ by columns of length $N$. It is possible to remove this ambiguity by removing all the columns of size $N$ until there are at most $N - 1$ rows.

Now we give some examples of representations of $U(N)$ and $SU(N)$ labelled by Young diagrams:
\begin{eqnarray}
	\begin{aligned}
		 & SU(3):\quad \ydiagram{1}\to \text{Fundamental}\quad \ydiagram{1, 1}\to \text{Antifundamental} \quad \ydiagram{2, 1}\to \text{Adjoint}          \\
		 & U(2): \quad \ydiagram{1}\to \text{Fundamental}\quad \ydiagram{2}\to \text{Sym}^2\text{Fundamental} \quad \ydiagram{1, 1}\to \text{Determinant}
	\end{aligned}
\end{eqnarray}

At various points in the text, we need to use the notation for the root lattice of $SU(N)$ and $U(N)$:
\begin{eqnarray}\label{eq:root_lat}
	\Lambda_R(SU(N)) = \Lambda_R(U(N)) = \{r = (r_1,..., r_N)\in \Z^N, \sum_{i= 1}^N r_i = 0\}
\end{eqnarray}
The inner product between two vectors is the usual Euclidean inner product, regardless of whether their components are integers. For example, $(r, r) = \sum_{i = 1}^N r_i^2$. In this convention, all the roots of $SU(N)$ (e.g. $(1, -1, 0,0,...)$) have norm squared equal to 2. A special vector known as the Weyl vector is defined as:
\begin{eqnarray}\label{eq:weyl_vector}
	\rho = (\frac{N - 1}{2}, \frac{N-3}{2}, \frac{N-5}{2}..., \frac{1 - N}{2})
\end{eqnarray}

Now we move onto the affine characters
\begin{eqnarray}
	\chi_{\lambda}^{\widehat{U(N)}_k}(q, x_1,..., x_N)\quad \chi_{\lambda}^{\widehat{SU(N)}_k}(q, x_1,..., x_N)
\end{eqnarray}
of $\widehat{U(N)}_k$ and $\widehat{SU(N)}_k$ respectively. $q$ is the fugacity for the zeroth Virasoro generator $L_0$. $x_i$ are the diagonal values of the Cartan torus of $U(N)$ (or $SU(N)$). In the $U(N)$ case, no constraint is placed on the fugacities $x_i$ except that they are nonzero complex numbers. In the $SU(N)$ case, we impose the constraint $\prod_{i = 1}^N x_i = 1$. $\lambda$ is an integrable highest weight at level $k$ in the sense that
\begin{eqnarray}
	\lambda_1\geq \lambda_2... \geq \lambda_N\quad \lambda_1 - \lambda_N\leq k
\end{eqnarray}
These conditions can be translated to the Young diagram of $\lambda$. In the $SU(N)$ case, if we assume $\lambda_N = 0$, the Young diagram must have at most $N - 1$ rows and $k$ columns. We use the convention that all affine characters start at $q^0$. An integrable weight of $SU(N)$ at level $k$ can be naturally extended to a representation of $\widehat{SU(N)}$ at level $k$ such that the zeroth-grade subspace of an affine representation equals the $SU(N)$ representation with highest weight $\lambda$. Therefore, we always label representations of $\widehat{SU(N)}_k$ using highest weights of $SU(N)$. A similar result holds for $U(N)$. The relationship between the two characters are as follows (when the product of the $x_i$ equals 1):
\begin{eqnarray}
	\chi_{\lambda}^{\widehat{U(N)}_k}(q, x_1,..., x_N) = (q;q)_\infty^{-1}\chi_{\lambda}^{\widehat{SU(N)}_k}(q, x_1,..., x_N) \quad \prod_{i = 1}^N x_i = 1
\end{eqnarray}
In other words, the only difference between the two characters is the $(q;q)_\infty^{-1}$ factor and an overall $U(1)$ charge, where the $U(1)$ embedding in $U(N)$ is given by $x_i = x, \forall i, x\in U(1)$. In this convention, we do not need to specify the $\widehat{U(1)}$ level of $\widehat{U(N)}_k$ as we do not include any $\widehat{U(1)}$ lattice sum in the $\widehat{U(N)}_k$ characters.

\section{Inner Products of Affine Characters}\label{app:inner_product}
Let $\chi^{\widehat{U(N)}_k}_\lambda(q, s_1,..., s_N)$ be a $\widehat{U(N)}_k$ character. We refer the reader to the previous appendix for our conventions. The aim of this appendix is to complete the proof of the diagonal factorization \eqref{eq:diagonal_fact} by showing that affine characters are orthogonal with respect to the following inner product \footnote{We thank Davide Gaiotto for pointing this out. See section 3.1 of \cite{Anempodistov2025} for a similar manipulation.}
\begin{eqnarray}\label{eq:inner_product_affine_ch}
	\begin{aligned}
		  & \frac{1}{N!}
		\oint_{|s_i| = 1} \prod_{i = 1}^N\frac{ds_i}{2\pi i s_i}
		\prod_{i\neq j}\left(1 - \frac{s_i}{s_j}\right)
		\prod_{i, j = 1}^N (q s_i / s_j;q)_\infty^2
		\chi_\lambda^{\widehat{U(N)}_k}(q, s)
		\chi_\mu^{\widehat{U(N)}_k}(q, s^{-1}) \\
		= &
		\delta_{\lambda \mu}\sum_{r \in \Lambda_R(U(N))}q^{(N + k) (r, r) + 2(r, \lambda + \rho)}
	\end{aligned}
\end{eqnarray}
where $\delta_{\lambda \mu} = 1$ when $\lambda = \mu$ and $0$ otherwise. $\rho$ is the Weyl vector \eqref{eq:weyl_vector} and $\Lambda_R(U(N))$ is the root lattice of $U(N)$ \eqref{eq:root_lat}. The contour is a direct product of unit circles. We have again written $s = (s_1,..., s_N), s^{-1} = (s_1^{-1},..., s_N^{-1})$. Our convention is that affine characters start at $q^0$ so the right-hand side also starts at $q^0$. When $q\to 0$, this identity reduces to the orthogonality of $U(N)$ characters with respect to the Haar measure. We refer the reader to \eqref{eq:ex_inner_prod_affine_ch} for an example of this inner product and appendix \ref{app:su(n)} for the convention of the inner products in the power of $q$. The proof of this formula consists of the following steps:

\textbf{STEP I:} We replace both characters using the Weyl character formula \cite{DiFrancesco}:
\begin{eqnarray}\label{eq:weyl_kac_ch}
	\chi_\lambda^{\widehat{U(N)}_k}(q, s) = \frac{\sum_{\hat{w}\in \widehat{W}} \det(\hat{w})\exp(\hat{w}(\hat{\lambda} + \hat{\rho}))}{\exp(\rho)\prod_{N\geq i > j\geq 1}(1 - s_i/s_j)\prod_{i, j = 1}^N\left(qs_i/s_j;q\right)_\infty}
\end{eqnarray}
The expression $\exp(\rho)$ stands for $\prod_{i = 1}^N s_i^{\rho_i}$ and appears in the denominator of the ordinary Weyl character formula for $U(N)$.
Now we explain what the numerator means. The vectors $\hat{\lambda} = (\lambda, k, 0) \in \R^{N + 2}$ and $\hat{\rho} = (\rho, N, 0) \in \R^{N + 2}$ are affine extensions of $\lambda$ and $\rho$. Their sum equals $\hat{\lambda} + \hat{\rho} = (\lambda + \rho, k + N, 0) \in \R^{N + 2}$. The affine Weyl group $\widehat{W}$ is the semidirect product Weyl group $W$ with the root lattice $\Lambda_R(U(N))$, the latter being the normal subgroup. We write an element $\hat{w}$ of $\widehat{W}$ as $wr$ where $w\in W, r\in \Lambda_R(U(N))$. The group law is $w  r = \tilde{r}  w$ where $\tilde{r} = w(r)$, the ordinary Weyl action on $r$. In the case of $U(N)$, the Weyl group $W$ is the symmetric group $S_N$ and it acts on $\R^N$ by permuting the components. The action of $wr$ (an element in the affine Weyl group $\widehat{W}$) on $\R^{N + 2}$ is given by \cite{DiFrancesco}
\begin{eqnarray}
	(w r)(\mu, m, n) = (w(\mu + mr), m, n + \frac{1}{2}m(r, r) + (\mu, r))
\end{eqnarray}
where $\mu \in \R^N, m\in \R, n\in\R$. Therefore, the action fixes the middle component $m$ and the value of $m$ tells us how it acts on the other components. Now we substitute in $\mu = \lambda + \rho, m = k + N,n = 0$. The result is
\begin{eqnarray}
	(wr)(\hat{\lambda} + \hat{\rho}) = (w(\lambda + \rho + (k + N)r), k + N, \frac{1}{2}(k + N)(r, r) + (\lambda + \rho, r))
\end{eqnarray}
The determinant $\det(\hat{w}) = \det(w)$ equals $1$ if $w$ is orientation preserving and $-1$ otherwise. For example, if $w$ is the reflection along a simple root, $\det(w) = -1$. We use $s = (s_1,..., s_N)$ to grade the first component $w(\lambda + \rho + (k + N)r)$ and $q$ to grade the last component $(1/2)(k + N)(r, r) + (\lambda + \rho, r)$. We do not assign any fugacity to the middle component $N + k$ as it depends only on $N$ and $k$. Therefore, the numerator can be written as:
\begin{eqnarray}
	\sum_{\hat{w}\in \widehat{W}} \det(\hat{w})\exp(\hat{w}(\hat{\lambda} + \hat{\rho})) = \sum_{\substack{w\in W\\ r\in \Lambda_R(U(N))}}\det(w)q^{(k + N)(r, r)/2 + (\lambda + \rho, r)} s^{w(\lambda + \rho + (k + N)r)}
\end{eqnarray}

\textbf{STEP II:} We rewrite the sum over the Weyl group in the previous step in the following way:
\begin{eqnarray}\label{eq:lambda_tilde}
	\sum_{w\in W}\det(w)s^{w(\lambda + \rho + (k + N)r)} = \pm \sum_{w\in W} \det(w) s^{w(\tilde{\lambda} + \rho)}
\end{eqnarray}
where $\tilde{\lambda} = (\tilde{\lambda}_1, \tilde{\lambda}_2,...)\in \R^N$ is a highest weight: $\tilde{\lambda}_1 \geq \tilde{\lambda}_2..$. This step is essential as the right hand side is (up to a sign) the numerator of the ordinary Weyl character formula for $U(N)$. It can be combined with part of the denominator of \eqref{eq:weyl_kac_ch} to form an ordinary character of $U(N)$. To prove the existence of $\tilde{\lambda}$, we will prove the existence of a $\tilde{w}\in W$ such that $\tilde{\lambda} +\rho= \tilde{w}(\lambda + \rho + (k + N) r)$. If we then substitute this formula for $\tilde{\lambda} + \rho$ into the right hand side and perform a change of variables $w\tilde{w} \to w$ in the sum, to deduce that the overall sign of the right hand side is $\det(\tilde{w})$. To prove the existence of $\tilde{w}$, we let $\tilde{w}$ to be the element in $W = S_N$ that permutes the components $\lambda + \rho + (k + N)r$ until they are decreasing. In other words, $\tilde{w}(\lambda + \rho + (k + N)r)$ lies in the fundamental chamber. Then we set $\tilde{\lambda} =\tilde{w}(\lambda + \rho + (k + N) r) - \rho$. We need to check that $\tilde{\lambda}$ satisfies the following conditions for it to be a highest weight:
\begin{enumerate}
	\item $\tilde{\lambda}_i$ are all integers. This follows from the fact that $\rho$ and $\lambda + \rho + (k + N) r$ are either both integers or both half integers.
	\item $\tilde{\lambda}$ lies in the fundamental chamber: $\tilde{\lambda}_1\geq \tilde{\lambda}_2\geq ...$. We will show that the components of $\tilde{w}(\lambda + \rho + (k + N)r)$ are strictly decreasing. The components of $\tilde{w}(\lambda + \rho + (k + N)r)$ are decreasing by the definition of $\tilde{w}$. Hence it suffices to show that the components of $\lambda + \rho + (k + N)r$ are pairwise distinct. The maximum difference between the components of $\lambda + \rho$ is
	      \begin{eqnarray}
		      \lambda_1 + \rho_1 - \lambda_N- \rho_N = \lambda_1 - \lambda_N + N - 1\leq N + k - 1
	      \end{eqnarray}
	      Therefore adding $(k + N)r$, where $r$ is in the root lattice, would not make any of the two components equal. The gap between adjacent elements of $\rho$ is 1, hence the components of $\tilde{\lambda} =\tilde{w}(\lambda + \rho + (k + N)r) - \rho $ are decreasing and $\tilde{\lambda}$ lies in the fundamental chamber.
\end{enumerate}

\textbf{STEP III:} Now we use the ordinary Weyl character formula for $U(N)$
\begin{eqnarray}
	\frac{
		\sum_{w\in W} \det(w) s^{w(\tilde{\lambda} + \rho)}
	}{
		\exp(\rho) \prod_{N\geq i> j \geq 1}(1 - s_i/s_j)
	} =  \chi_{\tilde{\lambda}}^{U(N)}(s)
\end{eqnarray}
to write
\begin{eqnarray}\label{eq:weyl_kac_ch_numerator}
	\frac{\sum_{\hat{w}\in \widehat{W}} \det(\hat{w})\exp(\hat{w}(\hat{\lambda} + \hat{\rho}))}{\exp(\rho)\prod_{i > j}(1 - s_i/s_j)} = \sum_{r\in \Lambda_R(U(N))}\pm q^{ (k + N)(r, r)/ 2 + (\lambda + \rho, r)} \chi_{\tilde{\lambda}}^{U(N)}(s)
\end{eqnarray}
We stress that $\tilde{\lambda}$ is a function of $\lambda,  k, N, r$ defined via \eqref{eq:lambda_tilde}:
\begin{eqnarray}
	\tilde{\lambda} +\rho= \tilde{w}(\lambda + \rho + (k + N) r)
\end{eqnarray}
for a suitable choice of $\tilde{w}$ and the sign before the power of $q$ is $\det(\tilde{w})$.

\textit{Example:} Consider $\widehat{SU(2)}_k$. We write an element of the root lattice as $r = (m, -m), m\in \Z$ and $\lambda = (j, -j), j \in \Z/2_{\geq 0}$ to match the notation of \cite{DavideDualBC}. We have $\rho = (1/2, -1/2)$ \eqref{eq:weyl_vector}. The power of $q$ becomes
\begin{eqnarray}
	(k + N)(r, r)/2 + (\lambda + \rho, r) = (k + 2)m^2 + (2j + 1)m
\end{eqnarray}
Now we split sum over $m\geq 0$ and $m < 0$. When $m\geq 0$,
\begin{eqnarray}
	\lambda + \rho + (k + N)r = (j + 1/2 + (k + 2)m, -j - 1/2 - (k + 2)m)
\end{eqnarray}
is already in the fundamental chamber so we set $\tilde{w} = 1$ and obtain $\tilde{\lambda} = (j + (k + 2)m, -j - (k + 2)m)$. When $m < 0$, we have $j + 1/2 + (k + 2)m < 0$ so we need to perform a Weyl reflection and set $\tilde{w} = -1$. Hence $\tilde{\lambda} = (-j- 1 - (k + 2)m, j + 1 + (k + 2)m)$ in this case. Now we perform a change of variable $m\to -m$ when $m < 0$ and \eqref{eq:weyl_kac_ch_numerator} becomes (assuming $s_1s_2 = 1$)
\begin{eqnarray}
	\begin{aligned}
		  & \chi_{(j, -j)}^{\widehat{SU(2)}_k}(q, s_1, s_2) \; (q;q)_\infty (q s_1 / s_2;q)_\infty (q s_2 / s_1;q)_\infty \\
		= & \sum_{m \geq 0}q^{(k + 2)m^2 + (2j + 1)m}\chi^{SU(2)}_{(j + (k + 2)m, -j - (k + 2)m)}(s_1, s_2)               \\
		  & - \sum_{m \geq 1}q^{(k + 2)m^2 - (2j + 1)m}\chi^{SU(2)}_{( - j - 1+ (k + 2)m, j  + 1  - (k + 2)m)}(s_1, s_2)
	\end{aligned}
\end{eqnarray}
which is exactly (7.6) of \cite{DavideDualBC}.

We are now ready to simplify the inner product \eqref{eq:inner_product_affine_ch}. The factor $\prod_{i, j = 1}^N\left(qs_i/s_j;q\right)_\infty^2$ in the integrand of \eqref{eq:inner_product_affine_ch}
cancels the corresponding factors in the denominators of the character formula \eqref{eq:weyl_kac_ch}. Therefore, the inner product \eqref{eq:inner_product_affine_ch} can be written as
\begin{eqnarray}\label{eq:inner_product_interm}
	\begin{aligned}
		\frac{1}{N!}
		\oint_{|s_i| = 1} \prod_{i = 1}^N\frac{ds_i}{2\pi i s_i}
		\prod_{i\neq j}\left(1 - \frac{s_i}{s_j}\right) & \sum_{r\in \Lambda_R(U(N))}\pm q^{ (k + N)(r, r)/ 2 + (\lambda + \rho, r)}
		\chi_{\tilde{\lambda}}^{U(N)}(s)                                                                                                                               \\
		                                                & \sum_{r'\in \Lambda_R(U(N))}\pm q^{ (k + N)(r', r')/ 2 + (\mu + \rho, r')} \chi_{\tilde{\mu}}^{U(N)}(s^{-1})
	\end{aligned}
\end{eqnarray}
where $\tilde{\mu}$ is related to $\mu$ in the same way as $\tilde{\lambda}$ is related to $\lambda$.

\textbf{STEP IV:}
Finally, we use the orthogonality of the $U(N)$ characters with respect to the Haar measure to perform the integral \eqref{eq:inner_product_interm}. There are two cases:
\begin{enumerate}
	\item $\lambda\neq \mu$. In this case, we will show that the inner product \eqref{eq:inner_product_interm} is zero. In other words, we will prove that $\tilde{\lambda} \neq \tilde{\mu}$ for any choice of $r$ and $r'$. This follows if we can show that
	      \begin{eqnarray}
		      w(\lambda + \rho + (k + N)r) - \rho \neq  w'(\mu + \rho + (k + N)r') - \rho
	      \end{eqnarray}
	      for any choice of $w, w', r, r'$ as long as both sides lie in the fundamental chamber. This statement follows if we can prove that
	      \begin{eqnarray}
		      \lambda + \rho - w(\mu + \rho) \neq (k + N)r\quad \forall r\in \Lambda_R(U(N)), w\in W
	      \end{eqnarray}
	      Assume that the following is true:
	      \begin{eqnarray}
		      \lambda_i  + \rho_i - w(\mu + \rho)_i = (k + N) r_i\quad r_i\in \Z\quad \sum_{i = 1}^N r_i = 0\quad w \in W
	      \end{eqnarray}
	      A nonzero element of the root lattice must have both positive and negative components: $r_i > 0, r_j < 0$ for some $i\neq j$. In the previous step we showed that the gap between any two components of $\lambda + \rho$ is at most $k + N - 1$. The same result holds for $\mu + \rho$ and $w(\mu + \rho)$. Therefore, the gap between any two components of $\lambda_i  + \rho_i - w(\mu + \rho)_i$ is at most $2(k + N - 1)$. However, the gap between $(k + N)r_i$ and $(k + N)r_j$ is at least $2(k + N) > 2(k  + N - 1)$ and we have a contradiction. The case $r = 0$ can also be ruled out as $\lambda + \rho\neq \mu + \rho$ and both of them are strictly dominant.
	\item $\lambda = \mu$:  we will show that the $U(N)$ characters $\chi^{U(N)}_{\tilde{\lambda}}$ are all distinct to facilitate the computation. In other words, we will prove that $\lambda + \rho + (k + N)r$ and $\lambda + \rho + (k + N)r'$ do not lie on the same Weyl orbit when $r\neq r'$. This follows from a similar argument as above. If $w(\lambda + \rho) - \lambda - \rho = (k + N)r$ for some $r\in \Lambda_R(U(N))$ and $w\in W$, the maximal difference between components of $w(\lambda + \rho) - \lambda - \rho$ is at most $2( k+ N - 1)$ which forces $r = 0$ and $w = 1$. We can now perform the integral of \eqref{eq:inner_product_interm} to complete the proof of \eqref{eq:inner_product_affine_ch}.
\end{enumerate}
Finally, we prove the following formula for the inner products of $\widehat{U(N)}_0$ characters which are just $1$.
\begin{equation}\label{eq:level_0_inner_prod}
	\begin{aligned}
		  & \frac{1}{N!}\oint_{|s_i| = 1}
		\prod_{i = 1}^N \frac{ds_i}{2\pi i s_i}
		\prod_{i\neq j} \left(1 - \frac{s_i}{s_j}\right)
		\prod_{i, j = 1}^N (q s_i / s_j; q)_\infty^2                           \\
		= & (q;q)^2_\infty\sum_{r \in \Lambda_R(U(N))} q^{N(r,r) + 2(\rho, r)}
	\end{aligned}
\end{equation}
To prove this, we use the Weyl denominator formula \cite{DiFrancesco}:
\begin{equation}\label{eq:weyl_denom}
	\begin{aligned}
		  & (q;q)_{\infty}^{-1}
		\prod_{i, j = 1}^N (q s_i / s_j; q)_\infty
		\\
		= & \frac{
			\sum_{\hat{w}\in \hat{W}}
			\det(\hat{w})
			\exp(\hat{w}(\hat{\rho}))
		}{
			\exp(\rho)\prod_{N\geq i > j \geq 1}(1 - s_i/s_j)
		}
		\\
		= & \sum_{\sum r_i = 0} \pm q^{N(r,r)/2 + (\rho, r)}\chi_{\kappa(r)}^{U(N)}(s)
	\end{aligned}
\end{equation}
for some $\kappa(r)$ depending on $r$. The $\pm$ sign also depends on $r$. The last step follows from setting $k = \lambda = 0$ in \eqref{eq:weyl_kac_ch_numerator}. The $\kappa(r)$ are all distinct by the same argument as before. Now \eqref{eq:weyl_denom} is invariant under $s \to s^{-1}$. So we write
\begin{equation}
\begin{aligned}
    \prod_{i, j = 1}^N (q s_i / s_j; q)_\infty^2 =& (q;q)_\infty^2 \sum_{\sum r_i = 0} \pm q^{N(r,r)/2 + (\rho, r)}\chi_{\kappa(r)}^{U(N)}(s)\\
    &\times \sum_{\sum r_i = 0} \pm q^{N(r,r)/2 + (\rho, r)}\chi_{\kappa(r)}^{U(N)}(s^{-1})
\end{aligned}
\end{equation}

 We can substitute this equation into \eqref{eq:level_0_inner_prod} and use the orthonormality of the $U(N)$ characters to finish the proof.
\section{Difference Equations of Affine Characters from Outer Automorphisms} \label{app:diff_eq}
In this appendix, we provide an alternative proof of the difference equation \eqref{eq:diff_eq} for affine characters using outer automorphisms.  We will use the following outer automorphism $\sigma$\footnote{This outer automorphism rotates the Dynkin diagram of $\widehat{SU(N)}$ by one node.} which acts on the Cartan subalgebra of $\widehat{SU(N)}$ in the following way \cite{DiFrancesco}:
\begin{eqnarray}\label{eq:sigma_action_cartan}
	\sigma(z_1, z_2,...,z_N)= (z_N, z_1, z_2, ..., z_{N - 1}) - z_N c \quad \sigma(c) =  c\quad \sum_{i} z_i = 0
\end{eqnarray}
where $(z_1,..., z_N)$ (the diagonal matrix with diagonal values $z_1,...,z_N$) lies in the Cartan subalgebra of $SU(N)$ and $c$ is the central element of $\widehat{SU(N)}$ which acts on a level $k$ representation as $k$. Strictly speaking, one should write $\sigma((z_1,..., z_N))$. However, we removed a set of brackets to improve readability and will do the same for the rest of this section. The induced action of the outer automorphism on the operator $L_0$ \footnote{Our $L_0$ differs from the zeroth Virasoro generator by a multiple of the centre $c$. We normalize $L_0$ such that its eigenvalues start at 0 for $L(\lambda)$ (see the paragraph below). The outerautomorphism on $L_0$ is computed by demanding that its commutators with the affine Lie algebra modes are preserved. The image $\sigma(L_0)$ is therefore unique up to a multiple of the centre $c$. The lack of uniqueness is not an issue in  later computations.} is given by
\begin{eqnarray}\label{eq:sigma_action_L0}
	\sigma(L_0) =  L_0  - \omega_1
\end{eqnarray}
where we have defined
\begin{eqnarray}
	\omega_1 = \left(\frac{N-1}{N}, -\frac{1}{N}, -\frac{1}{N},..., -\frac{1}{N}\right)
\end{eqnarray}
an element of the Cartan subalgebra of $SU(N)$. We use the convention that the commutator of $L_0$ and all the negative modes of the affine lie algebra have positive eigenvalues. Therefore, the eigenvalues of $L_0$ are nonnegative on a integrable representation and starts at $0$.

Let $\lambda$ be an integrable representation of $SU(N)$ at level $k$ (see appendix \ref{app:su(n)} for more information). Precomposing the representation $\lambda$ (now regarded as a representation of $\widehat{SU(N)}_k$) with the automorphism $\sigma$ gives us another representation $\sigma(\lambda)$. For example, if $\lambda$ is the $k$th symmetric power of the fundamental representation, $\sigma(\lambda)$ is the vacuum representation. Written in equations, we have a composition of two maps:
\begin{eqnarray}
	\widehat{SU(N)}\xrightarrow{\quad\sigma\quad} \widehat{SU(N)} \xrightarrow{\quad\rho\quad} \text{End}(L(\lambda))
\end{eqnarray}
where $L(\lambda)$ is the representation space of $\lambda$ and $\rho$ is the representation. $\text{End}$ denotes the endomorphism algebra regarded as a Lie algebra. Our convention is that the eigenvalues of $\rho(L_0)$ on $L(\lambda)$ starts at $0$ and are nonnegative. The proof consists of the following steps:

\textbf{STEP I:} We compute the character of $\rho \circ \sigma$ in two different ways:
\begin{eqnarray}
	\begin{aligned}
		  & \tr_{L(\lambda)}\left(
		q^{\rho(\sigma(L_0))}
		\exp(2\pi i\rho(\sigma(z_1,..., z_{N - 1}, z_N)) )\right)                             \\
		= & \tr_{L(\lambda)} \left(
		q^{\rho(L_0 - \omega_1)}
		\exp(2\pi i \rho( z_N, z_1,..., z_{N - 1}))
		\right) \exp(-2\pi i k z_N)                                                           \\
		= & q^{-\lambda(\omega_1)}\chi_{\sigma(\lambda)}^{\widehat{SU(N)}_k}(q, x_1,..., x_N)
	\end{aligned}
\end{eqnarray}
where we have set $x_i = \exp(2\pi i z_i)$ to be the exponentiated fugacities.
For the first equality, we substitute in the action of $\sigma$ \eqref{eq:sigma_action_cartan}\eqref{eq:sigma_action_L0}. For the second equality, we use the fact that $\rho \circ \sigma$ is the representation $L(\sigma(\lambda))$ with the eigenvalue of $L_0$ shifted by $-\lambda(\omega_1)$ since the eigenvalue of $\rho(\sigma(L_0))$ starts at $-\lambda(\omega_1)$. We also remind ourselves of the convention that affine characters start at $q^0$. Now we write the second line using the character of $L(\lambda)$. The equality of the second line and the third line becomes
\begin{eqnarray}\label{eq:intermediate_diff_eq}
	\chi_\lambda^{\widehat{SU(N)}_k}(q, q^{1/N}x_1,..., q^{1/N}x_{N - 1}, q^{-(N - 1)/N}x_N)x_N^{-k} = q^{-\lambda(\omega_1)}\chi_{\sigma(\lambda)}^{\widehat{SU(N)}_k}(q, x_1,..., x_N)
\end{eqnarray}

As an example, take $\widehat{SU(2)}_1$ and set $x = x_1 = x_2^{-1}$. The weight $(1/2 ,-1/2)$ corresponds to the fundamental representation and $(0,0)$ corresponds to the vacuum representation. The characters are
\begin{eqnarray*}
	\chi_{(1/2, -1/2)}^{\widehat{SU(2)}_1}(q, x) = \frac{\sum_{n\in \Z} q^{n^2 + n}x^{2 n + 1}}{(q;q)_\infty}\quad \text{and} \quad \chi_{(0, 0)}^{\widehat{SU(2)}_1}(q, x) = \frac{\sum_{n\in \Z} q^{n^2}x^{2 n}}{(q;q)_\infty}.
\end{eqnarray*}
We set
\begin{align*}
	 & \lambda = (1/2, -1/2)\quad \sigma(\lambda) = (0, 0)\quad \omega_1 = (1/2, -1/2) \\ &\lambda(\omega_1) = 1/2 \times 1/2 + (-1/2) \times (-1/2) = 1/2
\end{align*}
One can check that
\begin{eqnarray*}
	\chi_{(1/2, -1/2)}^{\widehat{SU(2)}_1}(q, q^{1/2}x)x = q^{-1/2}\chi_{(0, 0)}^{\widehat{SU(2)}_1}(q, x)
\end{eqnarray*}
since
\begin{eqnarray}
	\sum_{n\in \Z} q^{n^2 + n}(q^{1/2}x)^{2n+ 1}x = q^{-1/2}\sum_{n\in \Z}q^{n^2 + 2n + 1}x^{2n + 2} = q^{-1/2}\sum_{n\in \Z}q^{n^2}x^{2n}
\end{eqnarray}

\textbf{STEP II:} Now we swap $x_i$ and $x_N$ in \eqref{eq:intermediate_diff_eq}:
\begin{eqnarray}
	x_i^{-k}\chi_\lambda^{\widehat{SU(N)}_k}(q, q^{1/N}x_1,.., q^{-N/(N - 1)}x_i,..., q^{1/N}x_N) = q^{-\lambda( \omega_1)}\chi_{\sigma(\lambda)}^{\widehat{SU(N)}_k}(q, x_1,..., x_N)
\end{eqnarray}
If we combine this equation and \eqref{eq:intermediate_diff_eq} and shift
\begin{eqnarray}
	x_1\to x_1 q^{-1/N},..., x_i\to x_i q^{(N - 1)/N},..., x_N\to x_N q^{-1/N}
\end{eqnarray}
We obtain the desired difference equation:
\begin{eqnarray}
	\chi_\lambda^{\widehat{SU(N)}_k}(q, x_1, ..., x_i  q, ..., x_N q^{-1})  = x_i^{-k} x_N^{k} q^{-k} \chi_\lambda^{\widehat{SU(N)}_k}(q, x_1, ..., x_N)
\end{eqnarray}
We can swap $x_N$ and $x_j$ to write it in the form \eqref{eq:diff_eq}.

\bibliographystyle{JHEP}
\bibliography{Interval_Pf}

\end{document}